\documentclass{aa}

\usepackage{txfonts}
\usepackage{graphicx}
\usepackage{natbib}
\usepackage{color}
\usepackage{euscript}
\usepackage{url}
\bibpunct{(}{)}{;}{a}{}{,} 

\begin{document}

\title{The long-wavelength emission of interstellar PAHs: characterizing the spinning dust contribution}
\author{Ysard N.
 \and Verstraete L.}
\offprints{N. Ysard, \email{nathalie.ysard@ias.u-psud.fr}}
\institute{Institut d'Astrophysique Spatiale, UMR8617, Universit\'e Paris-Sud, F-91405, Orsay, France}
\abstract
{The emission of cold dust grains at long wavelengths will soon be observed by the Planck and Herschel satellites and will provide new constraints on the nature of 
interstellar dust. In particular, the microwave galactic anomalous foreground detected between 10 to 90 GHz, proposed as coming from small spinning grains 
(PAHs), should help to define these species better. Moreover, understanding the fluctuations of the anomalous foreground quantitatively over the sky is crucial for 
CMB studies.}
{We focus on the long-wavelength emission of interstellar PAHs in their vibrational and rotational transitions. We present here the first model that coherently describes 
the PAH emission from the near-IR to microwave range.}  
{\small{We take quantum effects into account to describe the rotation of PAHs and compare our results to current models of spinning dust to assess the validity of the classical 
treatment used. Between absorptions of stellar photons, we followed the rovibrational radiative cascade of PAHs. We used the exact-statistical method of Draine \& Li to derive the 
distribution of PAH internal energy and followed a quantum approach for the rotational excitation induced by vibrational (IR) transitions. We also examined the influence 
of the vibrational relaxation scheme and of the low-energy cross-section on the PAH emission. We study the emissivity of spinning PAHs in a variety of physical conditions 
(radiation field intensity and gas density)}, search for specific signatures in this emission that can be looked for observationally,  and discuss how the anomalous foreground may 
constrain the PAH size distribution.}
{Simultaneously predicting the vibrational and rotational emission of PAHs, our model can explain the observed emission of the Perseus molecular cloud from the 
IR to the microwave range with plausible PAH properties. We show that for $\lambda \geq 3$ mm the PAH vibrational emission no longer scales with the radiation 
field intensity ($G_0$), unlike the mid-IR part of the spectrum (which scales with $G_0$). This emission represents less than 10\% of the total dust emission at 100 GHz. 
Similarly, we find the broadband emissivity of spinning PAHs per carbon atom to be rather constant for $G_0\leq 100$ and for proton densities $n_H<100$ cm$^{-3}$. 
In the diffuse ISM, photon exchange and gas-grain interactions play comparable roles in exciting the rotation of PAHs, and the emissivity of spinning PAHs 
is dominated by the contribution of small species (bearing less than 100 C atoms). We show that the classical description of rotation 
used in previous works is a good approximation and that unknowns in the vibrational relaxation scheme and low-energy cross-section affect the PAH rotational emissivity around 
30 GHz by less than 15\%.}
{The contrasted behaviour of the PAH vibrational and rotational emissivities with $G_0$ provides a clear prediction that can be tested against observations of anomalous and dust 
mid-IR emissions: this is the subject of a companion paper. \small{Comparison of these emissions complemented with radio observations (21 cm or continuum) 
will provide constraints on the fraction of small species and the electric dipole moment of interstellar PAHs. }}
\authorrunning{Ysard N. $\&$ Verstraete L.}
\titlerunning{Long-wavelength emission of interstellar PAHs}
\maketitle

\section{Introduction}

The mid-IR spectrum of the interstellar medium (ISM) shows prominent bands from 3 to 17~$\mu$m, which account for one third of the energy emitted by interstellar dust. Such 
bands are emitted by very small (subnanometric-sized) dust particles during internal energy fluctuations triggered by the absorption of a stellar photon (Sellgren 1984). 
The positions of these bands suggest there are aromatic, hydrogenated cycles in these grains. \citet{Leger_Puget} and \citet{Allamandola1985} proposed polycyclic aromatic 
hydrocarbons (PAHs) as the carriers of these bands. Despite two decades of experimental and theoretical efforts, the match (band positions and intensities) between data only 
available on small species and observations still remains elusive, as illustrated recently by the work of \citet{Peeters2002} and \citet{Kim}. Given their important role in the 
ISM (e.g., gas heating, Habart et al. 2001, UV extinction, Joblin et al. 1992), it is necessary to find other ways to constrain the properties of interstellar PAHs.

\noindent The Planck and Herschel data will soon trace the emission of cold interstellar grains in the interstellar medium. Due to their small size, PAHs are heated sporadically 
(every few months in the diffuse ISM) by absorption of stellar photons and have a high probability of being in low-energy states \citep{Li}. Interstellar PAHs may thus 
contribute significantly to the emission at long wavelength ($\lambda > 1$ mm). In this context, an unexpected emission excess called {\em anomalous foreground}, 
correlated to dust emission, has been discovered between 10 and 90 GHz \citep{Leitch, Oliveira2002}. In this spectral range, several galactic emission components (synchrotron, 
free-free, and thermal dust) contribute with a comparable magnitude, and only recently has the anomalous foreground been separated in WMAP data 
\citep{Miville2008}. Spinning, small dust grains were first proposed by \citet{DL98} (hereafter DL98) as a possible origin to this anomalous component. Since then, analysis of 
observations has suggested that the anomalous foreground is correlated to small grain emission \citep{Lagache, Casassus2006}.

\noindent In this paper, we study the emission of PAHs with particular emphasis on the long-wavelength part of the spectrum. In this spectral range, the emission is dominated by 
species in low-energy states for which each photon exchange represents a large energy fluctuation. To derive the internal energy distribution of PAHs, we use here the exact 
statistical method described in \citet{Li}. We include low-frequency bands to the vibrational mode spectrum of interstellar PAHs and examine the influence of the internal 
vibrational redistribution hypothesis (see Sect. \ref{p_de_e}) often used to describe their vibrational relaxation. We describe the rotation of PAHs with a quantum approach where 
specific processes in the rovibrational relaxation are naturally included (molecular recoil after photon emission; rovibrational transitions that leave the angular momentum of 
the molecule unchanged or $Q$ bands). We present in a variety of interstellar phases the rovibrational (IR) and broadband rotational (spinning) emission of PAHs in a consistent 
fashion. Our model results are compared to observations of the interstellar emission from the IR to the microwave range and show how the size distribution and electric dipole moments 
of interstellar PAHs may be constrained.

\noindent The paper is organized as follows. Section 2 describes the properties of PAHs adopted in this work. Section 3 discusses the internal energy distribution of PAHs and 
associated rovibrational emission. Sections 4 and 5 discuss the rotational excitation and emission of PAHs. In Sect. 6, we apply our model to the case of a molecular cloud 
in the Perseus arm. Finally, conclusions and observational perspectives are given in Sect. 7. 

\section{The properties of interstellar PAHs}

In the ISM, the stablest PAHs are found to be compact species \citep{Leger,LePage}. Small PAHs are planar, but above some poorly known size threshold (40 to 100 C atoms), 
interstellar formation routes may favour three-dimensional species (bowl- or cage-shaped) containing pentagonal cycles (see Moutou et al. 2000 and references therein). 
As we see later, small PAHs (containing less than 100 C atoms) dominate the rotational emission, so we assume that interstellar PAHs are planar with a hexagonal ($D_{6h}$) 
symmetry. The PAH radius is thus $a(\mathring{A}) = 0.9\sqrt{N_{C}}$, where $N_{C}$ is the number of carbon atoms in the grain \citep{Omont}. The formula for such molecules 
is $C_{6p^{2}}H_{6p}$, and their hydrogen-to-carbon ratio is $\frac{H}{C} = f_{\rm H} \sqrt{\frac{6}{N_{C}}}$, where $f_{\rm H}$ is the hydrogenation fraction of PAHs. In 
this work, we assume that PAHs are fully hydrogenated ($f_H=1$).

\subsection{Absorption cross-section}
\label{abs_cross_sect}

The excitation, cooling, and emission of PAHs depend on their absorption cross-section (Sect. \ref{p_de_e}), which we describe now. We took the visible-UV cross-section from 
\citet{Verstraete92} and applied their size-dependent cut-off for electronic transitions in the visible-NIR range. The resulting cross-section compares well to the available data 
\citep{Joblin1992}. In Appendix \ref{appendix_sigma}, we discuss the mid-IR bands considered in this work.
Each vibrational mode is assumed to be harmonic, and the corresponding band profile has a Drude shape \citep{Li}. The width $\Delta\nu$ is inferred from astronomical 
spectra, and the peak value $\sigma$ is chosen so that the integrated cross-section $\sigma\Delta\nu$ is equal to the value measured in the laboratory. By adopting the observed bandwidth, 
we empirically account for the complex molecular relaxation and band broadening \citep{Pech2002,Mulas2006} in interstellar PAHs. We adopt here the band strengths given in \citet{Pech2002} 
that were inferred from laboratory data. Different definitions of the PAH IR cross-sections have been proposed by \citet{Rapacioli2005}, \citet{Flagey2006}, and Draine \& Li (2007). 

\begin{figure}[!t]
\resizebox{\hsize}{!}{\includegraphics{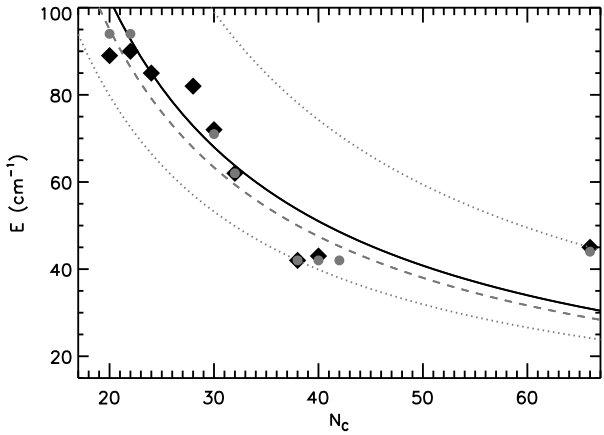}}
\caption{Frequencies of the first (lowest energy) vibrational mode versus $N_C$ for PAH cations 
(black diamonds) and PAH neutrals (grey circles) from the Mallocci database\protect\footnotemark (see text). The solid line is the relationship we adopt between the band 
position of cations and $N_C$, and the dashed line shows the case of neutrals. The dotted lines show extreme cases for this relationship.}
\label{E_weight}
\end{figure}

\noindent We used the database of \citet{Malloci2007} to define an average broadband cross-section of the far-IR vibrations of PAHs. At frequencies below 500 cm$^{-1}$, each 
species features many modes. However, for compact species, modes accumulate in three definite frequency ranges: modes with a frequency of less than 100~cm$^{-1}$, modes between 100 
and 200~cm$^{-1}$, and modes between 200 and 500~cm$^{-1}$. We therefore model the far-IR cross-section of compact PAHs with 3 modes (Table \ref{far_IR}). The frequency of each 
mode is the average of all modes falling within the given energy range, weighted by their corresponding integrated cross-sections. We find that the frequency of the lowest energy mode 
depends on the molecular size as $N_C^{-1}$ (see Fig. \ref{E_weight}). Conversely, for the two other modes, the average energy is instead independent of the size. The integrated cross-section 
for these 3 modes was estimated as follows. From the Mallocci database, we first derived the fraction of $\sigma\Delta\nu$ for each of the 3 modes. Then, we assumed that the total $\sigma
\Delta\nu$ below 500~cm$^{-1}$ is given by the integral of the absorption cross-section of \citet{Schutte}: $\sigma_{\rm FIR}=4.3\times 10^{-20}\;\lambda^{-1.24}$ cm$^2$ per 
C-atom. The integrated cross-sections of each of the 3 modes was finally obtained by multiplying the former value by the $\sigma\Delta\nu$ fractions inferred from the database. 

\begin{figure}[!t]
\resizebox{\hsize}{!}{\includegraphics{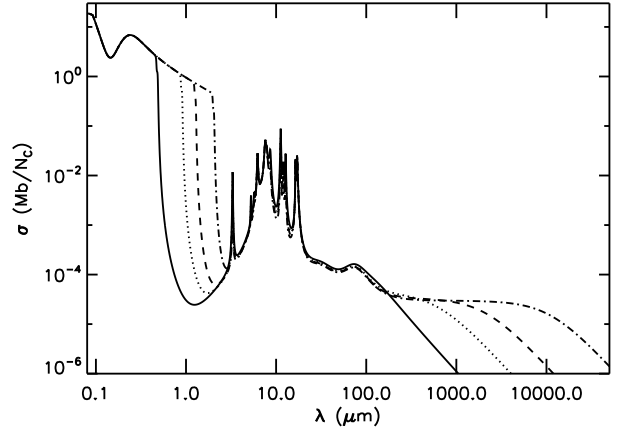}}
\caption{Absorption cross-section of PAH cations per carbon atom (1~Mb~=~10$^{-18}$~cm$^{2}$). The solid, dotted, dashed, and 
dot-dashed lines show the cases for $N_{C}$~=~24, 54, 96, and 216, respectively.}
\label{sigma}
\end{figure}

\noindent The parameters of the far-IR bands adopted here are given in Table \ref{far_IR}, and the full cross-section is displayed in Fig. \ref{sigma}. We show in Sect. \ref{p_de_e} 
that this set of IR bands provides a good match to observed interstellar spectra. Finally, we note that the in-plane or out-of-plane character of each vibrational band is important because 
of the different associated weights in the rotational excitation (Sect. \ref{rovib}). This character is indicated in Tables \ref{mid-IR} for the mid-IR bands \citep{Socrates}. In the case of 
the far-IR bands, this character is not as well known and we assume that 1/3 (2/3) of the oscillator strength come from out-of-plane (in-plane) transitions
respectively (Table \ref{far_IR}). 

\begin{table}[t]
\begin{minipage}{0.5\textwidth}
\centering
\caption{Far-IR rovibrational bands of PAH cations adopted in this work, with the percentage of the total oscillator strength in each band.}
\label{far_IR}
\centering
\begin{tabular}{cccccc}
\hline
\hline
$\lambda_{cations}$\footnote{Similar bands were obtained for PAH neutrals.} & $\nu_{cations}$ & $\Delta \nu$ & $\sigma_{i}/N_C$  & $\%$  & Type\footnote{In-plane (ip) or out-of-plane (op) character of the band.} \\
($\mu$m)            & (cm$^{-1}$)     & (cm$^{-1}$)  & $(\rm{10^{-20}cm^{2}})$ & (cations)        \\
\hline
30.2                & 331             & 300          & 9.6$\times 10^{-3}$       & 69.5      & 2/3 ip   \\
                      &                    &                 &                                      &              & 1/3 op \\
74.1                & 135             & 100          & 9.9$\times 10^{-3}$       & 23.9      & 2/3 ip \\
                      &                    &                 &                                      &              & 1/3 op \\
4.9$\times N_{C}$   & 2040/${N_{C}}$  & 100          & 2.7$\times 10^{-3}$       & 6.6       & 2/3 ip \\
                      &                    &                 &                                      &              & 1/3 op \\
\hline
\end{tabular}
\end{minipage}
\end{table}

\footnotetext{The molecules considered are perylene ($C_{20}H_{12}$), benzo[g,h,i]perylene ($C_{22}H_{12}$), coronene ($C_{24}H_{12}$), bisanthene ($C_{28}H_{14}$), 
dibenzo[bc,ef]coronene ($C_{30}H_{14}$), ovalene ($C_{32}H_{14}$), circumbiphenyl ($C_{38}H_{16}$), circumanthracene ($C_{40}H_{16}$), cirumpyrene ($C_{42}H_{16}$), 
and circumovalene ($C_{66}H_{20}$).}

\subsection{The rigid rotor model}
\label{pah_prop}

\begin{figure}[!t]
\centering
\resizebox{\hsize}{!}{\includegraphics{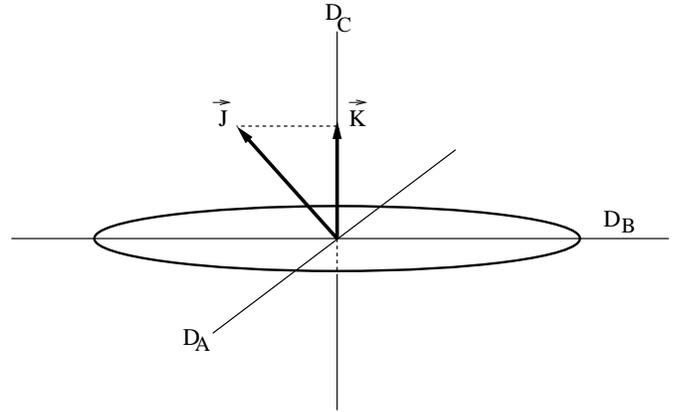}}
\caption{Symmetric top molecule: $D_{A}$, $D_{B}$, and $D_{C}$ are the principal axis of inertia and $J$ the total angular 
momentum of the molecule with $K$ its projection along $D_{C}$.}
\label{PAHs}
\end{figure}

\noindent While describing the rotation of a molecule, the relevant operator is the total angular momentum $\overrightarrow{J}$, which includes the electrons 
and nuclei contributions without the spin. We note that $D_{A}$, $D_{B}$, and $D_{C}$ are the principal axis of inertia. We assumed that PAHs are oblate symmetric top 
molecules with the axis $D_{C}$ perpendicular to the plane of the molecule and parallel to $Oz$. We called $I_{C}$ the inertia moment with respect to $D_{C}$ 
and $I_{A}$, $I_{B}$ the inertia moments with respect to $D_{A}$ and $D_{B}$, which were taken to be parallel to $Ox$ and $Oy$. The rotational Hamiltonian 
is then
\begin{equation}
H = \frac{J_{x}^{2}}{2I_{A}}+\frac{J_{y}^{2}}{2I_{B}}+\frac{J_{z}^{2}}{2I_{C}}
\end{equation}
where $J_{x}$, $J_{y}$, and $J_{z}$ are the projections of $\overrightarrow{J}$ along the three inertia axis. Given the large number of carbon atoms 
in interstellar PAHs, we assume here that they are uniform disks with $I_{A} = I_{B} = \frac{Ma^{2}}{4} = \frac{I_{C}}{2}$ (symmetric tops), where $M$ 
is the molecular mass. With $D_{C} \parallel Oz$, we get
\begin{equation}
H = \frac{J^{2}}{2I_{B}} + J_{z}^{2}\left(\frac{1}{2I_{C}}-\frac{1}{2I_{B}}\right) \; ,
\end{equation}
and the rotational energy is
\begin{equation}
\label{e_rot}
E_{rot} = BJ(J+1)+(C-B)K^{2}
\end{equation}
where $B/hc = \frac{\hbar}{4\pi cI_{B}} = 7 \ N_{C}^{-2}$ cm$^{-1}$ (neglecting the contributions of H to the molecular mass) and $C = B/2$ are the 
rotational constants associated to $D_{B}$ and $D_{C}$. The quantum number $K$ is the absolute value of the $J_{z}$-eigenvalues. For a symmetric top 
molecule, the selection rules for rovibrational electric dipole transitions are $\Delta J = 0, \pm 1$ and $\Delta K = 0, \pm 1$. By defining the dipole moment 
as $\overrightarrow{\mu} = \overrightarrow{\mu_{z}} + \overrightarrow{\mu_{B}}$, with $\overrightarrow{\mu_{z}}$ along $D_{C}$ and $\overrightarrow{\mu_{B}}$ 
in the molecular plane along $D_{B}$ \citep{Townes}, two kinds of transitions can be distinguished. First, the {\it parallel} transitions with $\Delta K = 0$ 
for which the change of dipole moment in the transition is parallel to the top axis ($D_C$) of the molecule. Second, the {\it perpendicular} transitions with 
$\Delta K = \pm 1$, for which the change of dipole moment is perpendicular to $D_C$. Parallel transitions thus correspond to {\em out-of-plane} 
vibrations, whereas perpendicular transitions correspond to {\em in-plane} vibrations.

\noindent Since the available microwave data are broadband observations ($\lambda/\Delta\lambda$ of the order of a few), we make several simplifying 
assumptions in the description of the rotational motion of PAHs. We thus assume that the rotational constant $B$ is the same in all vibrational levels, and within the 
framework of a rigid rotor model, we neglect the centrifugal distortion terms in the energy equation that are usually small for large molecules \citep{Herzberg,Lovas}.

\subsection{Electric dipole moment}

The rotational emission of PAHs depends on their permanent electric dipole moment, $\mu$. Symmetric (D$_{6h}$), neutral, and fully 
hydrogenated PAHs have $\mu\sim 0$. Spectroscopic analysis of their IR emission bands suggests that interstellar PAHs can hqve a 
cationic form that is partially hydrogenated \citep{LePage}, and maybe also substituted \citep{Peeters2002,Peeters04}. For instance, a PAH 
having lost one H atom has $\mu\sim$ 0.8 to 1 D, and a PAH cation where a C atom has been substituted by N would also have $\mu\sim$ 
0.1 to 1.5 D depending on its size (T. Pino, private communication). Moreover, it has been proposed that non-planar PAHs containing pentagonal 
rings may exist in the ISM (see Moutou et al. 2000 and references therein): such species are known to have large dipole moments as recently 
measured on coranulene, C$_{20}H_{10}$, $\mu = 2$ D \citep{Lovas}. In this work, we express the electric dipole moment of interstellar PAHs 
as in DL98:
\begin{equation}
\label{equation_mu}
\mu (N_C) = m\sqrt{N_{at}} + 4.3\times 10^{-2} \sqrt{N_C} Z \simeq m\sqrt{N_{at}} \;\;\;\; {\rm (Debye)}
\end{equation}
where $N_{at}$ is the total atoms number in the molecule, $Z$ is its charge, and $m$ a constant. Unless otherwise stated, we will use $m=0.4$ D.

\section{Internal energy distribution and rovibrational IR emission of isolated interstellar PAHs}
\label{p_de_e}

\begin{figure}[!t]
\centering
\resizebox{\hsize}{!}{\includegraphics[angle=90]{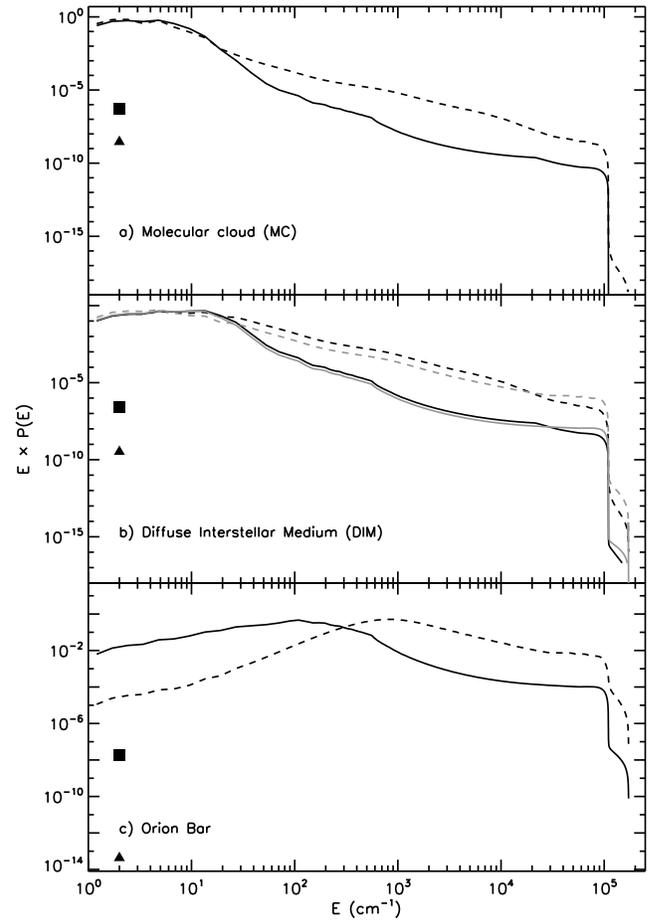}}
\caption{Internal energy distribution of PAHs with $N_{C} = 24$ (solid line and box for $P(0)$) and $N_{C} = 216$ 
(dashed line and triangle) in the case of $a)$ an MC ($G_0=10^{-2}$), $b)$ the diffuse ISM ($G_{0} 
= 1$) and $c)$ the Orion Bar (the radiation field is the sum of the CMB, the ISRF and a blackbody at $37\,000$ K corresponding 
to $G_{0} = 14,000$). To illustrate the effect of the radiation field hardness, we show in $b)$ the case of an Orion Bar type 
radiation field scaled down to $G_0=1$ (grey lines).}
\label{P_E}
\end{figure}

\noindent After the absorption of a visible-UV photon, a PAH cools off by emitting IR rovibrational photons. These photons reduce its angular momentum and also 
may increase it by recoil (a purely quantum effect, see Sect. \ref{rovib}). Previous studies have mostly used a thermal description of molecular 
cooling. In fact, since PAHs spend a large fraction of their time at low internal energies (see Fig. \ref{P_E}), their emission at long wavelengths cannot 
be considered as a negligible fluctuation compared to their total internal energy. While estimating the emission of PAHs at low temperatures, the validity 
of a thermal approach is questionable. Because of the rapid energy redistribution between interactions (photon exchange or collision with gas phase 
species), PAHs rapidly reach thermodynamical equilibrium while isolated \citep{Leger}. To describe this situation, we used the exact-statistical method 
of \citet{Li} to derive the stationary internal energy distribution of PAHs.

\noindent The energy distribution, $P(E)$, depends on (a) the energy density of the exciting 
radiation field $u_{E}=4\pi\nu I_{\nu}$ (where $I_{\nu}$ is the brightness), (b) the absorption cross-section, $\sigma$, of interstellar PAHs, and (c) their 
rovibrational density of states, $\rho(E)$. For the stellar contribution, we used the local interstellar radiation field (ISRF, Mathis et al. 1983) or a blackbody. 
We sometimes scale this stellar radiation field with a factor, $G_{0}$, equal to 1 in the case of the Mathis field
\footnote{$G_{0}$ scales the radiation field intensity integrated between 6 and 13.6 eV. The Mathis radiation field, $G_0=1$, corresponds to an intensity of $1.6\times 
10^{-3}$ erg/s/cm$^2$.}. Throughout this work, the contribution of the CMB is included in the radiation field. The vibrational absorption cross-section used has 
been described previously, and for simplicity, we did not include rotational bands. Unless otherwise stated, we took the cross-section of PAH cations. The density of 
states was obtained by first deriving the vibrational mode spectrum from a Debye model and then applying the algorithm of \citet{Beyer} for each molecular size (see Appendix \ref{modes}).

\begin{figure*}[!ht]
\centerline{
\begin{tabular}{cc}
\includegraphics[width=9cm]{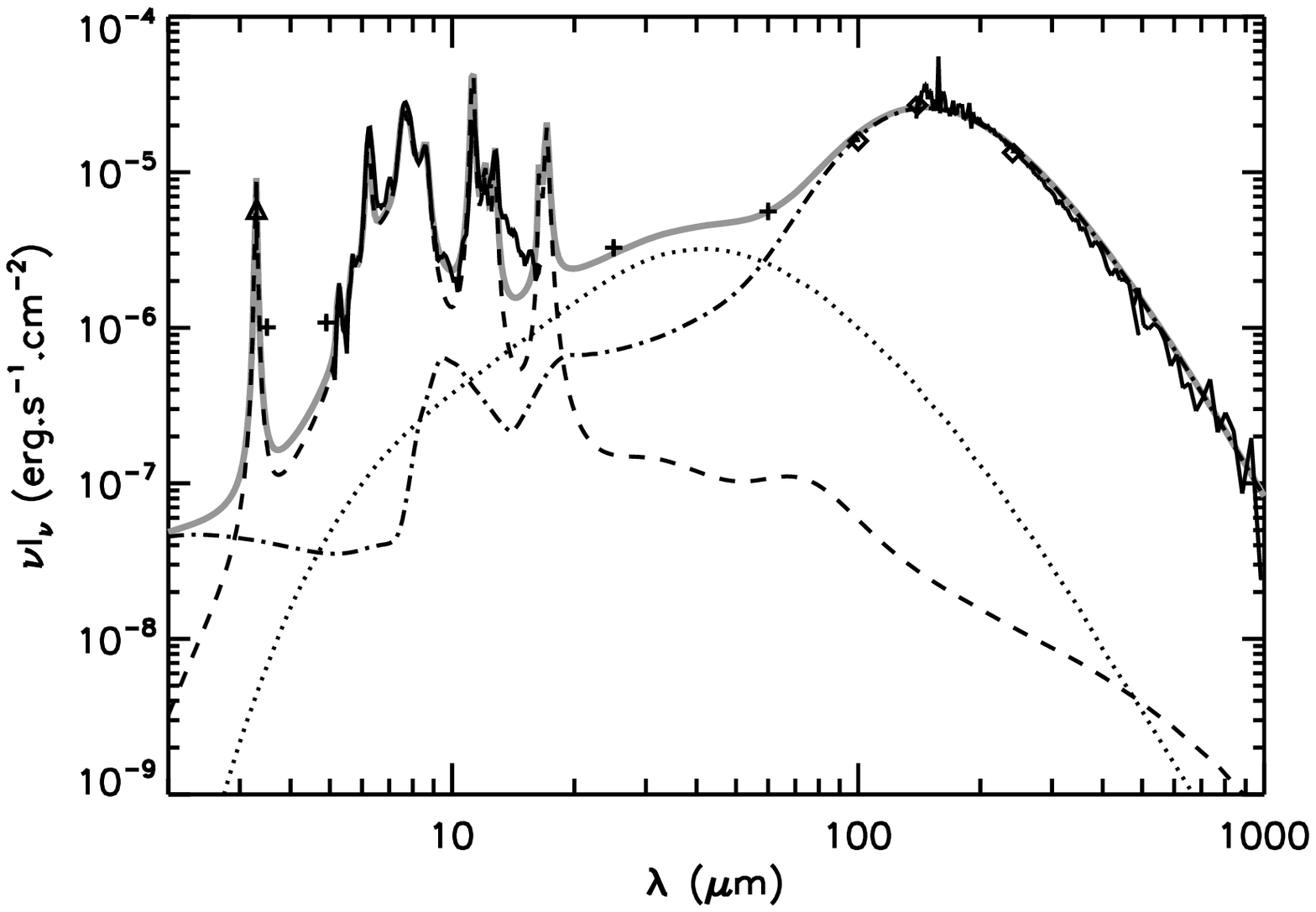} & \includegraphics[width=9cm]{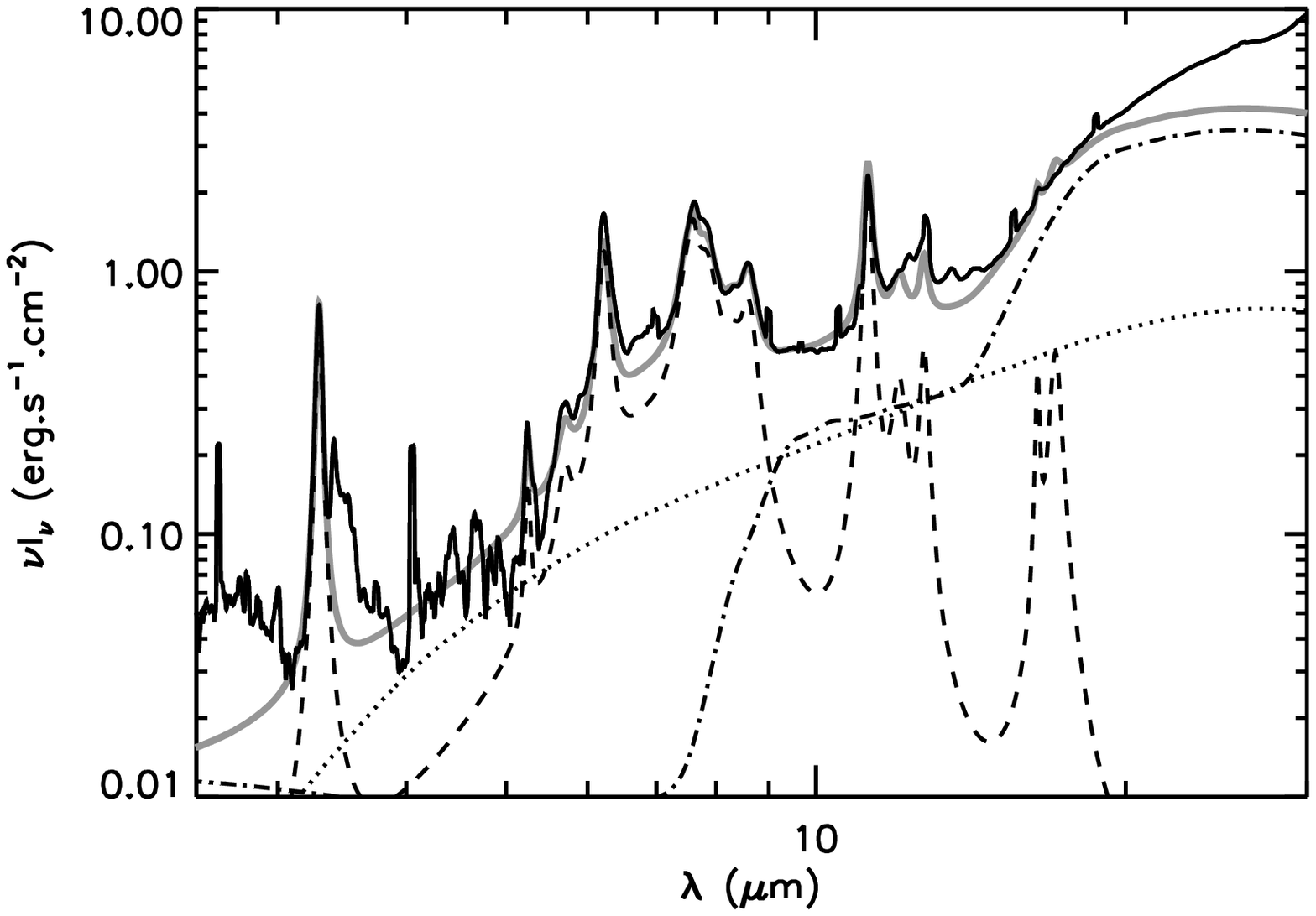}\\
\end{tabular}}
\caption{{\em Left:} Dust IR emission of the DIM (solid black lines and symbols) with $N_{H} \sim 10^{20}$~cm$^{-2}$ \citep{Boulanger2000}. The model dust emission is overlaid 
in bold grey. MRN type size distributions are used for all dust populations. The PAH vibrational contribution (dashed line) is from the present model with $N_{C} = [18,96]$, a 1:3 
mixture of neutrals and cations, full hydrogen coverage, and 43 ppm of carbon. The contributions of larger grains are from the model described in \citet{Compiegne2008}. The dotted 
line is the contribution of graphitic very small grains of radii $a=0.9$ to 4 nm containing 39 ppm of C. The dot-dashed line is the contribution of silicate big grains with $a=0.4$ to 
250 nm using 37 ppm of Si. {\em Right:} Dust emission from the Orion Bar (the noisy black line is the ISO-SWS spectrum) with $N_{H} \sim 1.8\times 10^{21}$~cm$^{-2}$. The model 
dust parameters are as above but for PAHs less abundant (18 ppm of C) and more ionized (82\% are cations).}
\label{emission_IR} 
\end{figure*}

\noindent After each absorption of a stellar photon, we assume that the excitation energy of the PAH is rapidly redistributed among all the vibrational 
modes \citep{Mulas2006}. This radiationless and isoenergetic process is called internal vibrational redistribution (IVR).
The energy distribution is then computed according to the exact-statistical formalism described in \citet{Li}. 
The energy scale (referenced to the zero point energy) is divided into bins of energy $E_{i}$ and width $\Delta E_{i}$ with $i = 0, 1, ..., M$ ($M = 500$) 
with $0 \leq E_i \leq 2E_L$ where $E_L=1.1 10^5$ cm$^{-1}$ is the Lyman limit. When its energy is below that of the first excited vibrational state, 
the molecule is in a rotational state. Letting $P_{i}$ be the probability of having a PAH in the energy bin $i$ we have
\begin{equation}
\frac{dP_{i}}{dt} = \sum_{j \ne i}^{M} \; T_{j \rightarrow i}P_{j} \; - \; \sum_{i \ne j}^{M} \; T_{i \rightarrow j}P_{i}
\end{equation}
where $T_{j \rightarrow i}$ is the transition rate from the state $j$ to the state $i$ for a molecule. We solve this equation in the stationary case. More 
details can be found in \citet{Li}. Figure \ref{P_E} shows the energy distribution for three different radiation fields, representative of a molecular cloud, the 
diffuse interstellar medium (DIM), and the Orion Bar. We see that the most probable energy increases when the radiation field intensity $G_{0}$ increases, as 
expected for PAHs whose cooling by IR emission is interrupted more frequently by absorption events. The sharp cut-off at $\sim 1.1\times 10^5$ cm$^{-1}$ stems from 
the Lyman limit of photon energies in neutral regions. We also observe a tail at higher energies that comes from to multiphoton events, i.e., absorption of 
a second photon while the PAH has not completely cooled off. This tail becomes more significant as absorption events are more frequent, i.e., for large PAHs 
or intense radiation fields. Conversely, we show in Fig. \ref{P_E}b that the radiation field hardness has little influence on $P(E)$. 
To estimate the effect of the rotational absorption that we neglected, we included a band centred at 1 cm$^{-1}$ of width 1 cm$^{-1}$ and corresponding to 
$J=150$ ($\sigma/N_C\sim 9\times 10^{-22}$ cm$^2$/C, see Sects. \ref{rot} and \ref{paragraph_contrib}): we found $P(E)$ to be twice as large between 1 and 10 cm$^{-1}$ 
and unchanged otherwise. This would change the emission around 1 cm at very low flux levels. The rotational excitation rates (Sect. \ref{rot}), which depend on the populations 
of excited vibrational levels, are unaffected by this hypothesis.

\begin{figure}[!t]
\centering
\resizebox{\hsize}{!}{\includegraphics{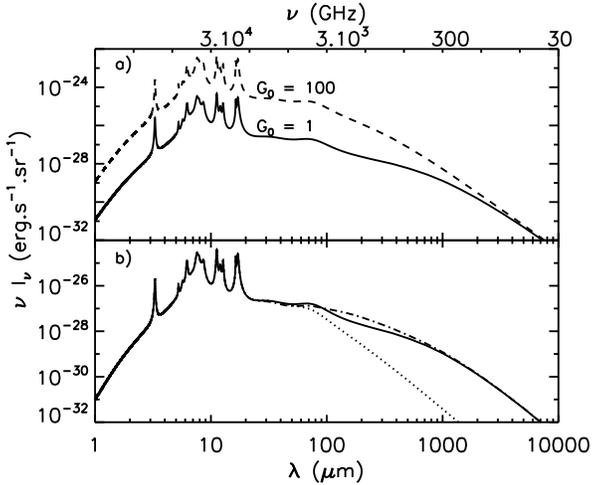}}
\caption{PAH rovibrational emissivity for a MRN size distribution with $N_C=18-216$. In a) we show the case of radiation field intensities corresponding to $G_0=1$ 
and 100 and in b) we illustrate the effect of IVR breakdown in case 1 (dotted line) and case 2 (dot-dashed line) (see section \ref{decouplage}).}
\label{IR_Go} 
\end{figure}

\noindent Knowing the internal energy distribution of PAHs, we can deduce their IR emission from the upper state $u$ by summing the contributions of all lower vibrational 
modes $l$ \citep{Li}:
\begin{eqnarray}
\nu F_{\nu} = \frac{2h\nu^{4}}{c^{2}} \; \sigma(\nu) \; P(h\nu) \; \left( 1 + \frac{u_{E}}{8\pi h^3 \nu^{3}} \right)
\end{eqnarray}
\begin{eqnarray}
\mathrm{where} \; P(h\nu) &=& \sum_{l} \; P_{l} \; \sum_{u=0}^{l-1} \; \frac{g_{u}}{g_{l}} \; \Delta E_{u} \; G_{ul}(h\nu) \; + \nonumber\\
& & \sum_{l} \; P_{l} \; \left(1 \; - \; \frac{h\nu}{\Delta E_{l}}\right)
\end{eqnarray}
is the number of rovibrational photons emitted at energy $h\nu$. The $G_{ul}$ functions are defined in \citet{Li}. The degeneracies $g_{u}$ and 
$g_{l}$ are the numbers of energy states in bins $u$ and $l$, respectively (see Appendix \ref{modes}). Figure \ref{emission_IR} shows a comparison 
between mid-IR observations and our model results with a power law size distribution $n_s(a)\sim a^{-3.5}$ (as in \citet{MRN} hereafter 
MRN) that provides a simple representation of the actual $n_s(a)$ \citep{Kim1994}. We see that the IR emission from the DIM is explained well with a standard abundance of carbon in 
PAHs and as large fraction of cations as in \citet{Flagey2006}. A similarly good match is obtained for the Orion Bar spectrum with strong PAH 
depletion probably reflecting an efficient photodestruction in this excited environment. Figure \ref{IR_Go}a shows the behaviour of the PAH vibrational emission  
with the radiation field intensity $G_0$: whereas the IR part scales with $G_0$, the mm-range ($\lambda>3$ mm) does not. The emission in this spectral range is produced 
mostly by PAHs in low-energy states for which the dominant heating source is absorption of CMB photons, and this is why this part of the spectrum does not vary with $G_0$. 
Finally, we emphasize that at low energies ($\lambda>20\;\mu$m), our model shows the broadband behaviour of the PAH vibrational emission. When a detailed vibrational mode spectrum
is used, the PAH emission at $\lambda >20 \; \mu$m is a superposition of numerous narrow bands that may be detectable with some of the instruments onboard Herschel \citep{Mulas2006}.

\subsection{Decoupling of vibrational modes}
\label{decouplage}

Our derivation of $P(E)$ assumes an efficient energy redistribution between vibrational modes (IVR) during the PAH relaxation. The IVR thus involves a coupling 
between vibrational modes via intramolecular transitions. However, it is known that IVR is no longer efficient when the excitation energy of the molecule is below some threshold 
$E_{dec}$ \citep{Mulas2006}, and we will then speak of IVR breakdown. When $E<E_{dec}$, the excitation of energetically accessible vibrational modes is frozen according 
to microcanonical statistics at energy $E_{dec}$. The molecule then continues to cool via emission of the excited vibrational modes.
To study the influence of decoupling on the long-wavelength emission, we consider two extreme cases for the relaxation at $E<E_{dec}$: 

\noindent - case 1: no IVR, cooling by IR forbidden modes. We assume these modes to be defined like those of Table \ref{far_IR} but with an oscillator strength $10^4$ lower and,

\noindent - case 2: no IVR, cooling only by the first vibrational mode at $2040/N_C$ (cm$^{-1}$) as defined in Table \ref{far_IR} 

\noindent and compare them to our model with no decoupling (IVR always fulfilled). We take $E_{dec}$ to be constant and equal to 500 cm$^{-1}$, a value relevant for PAHs 
containing 20 to 30 carbon atoms \citep{Joblin2008}. Actually, $E_{dec}$ is expected to decrease with molecular size because of the increasing density of states at 
a given energy \citep{Mulas1998}: the former cases therefore provide an upper limit to the variations in the PAH long-wavelength emissivity. Figure \ref{IR_Go}b shows 
that, as expected, case 1 provides a lower limit to the long-wavelength emission of PAHs, whereas case 2 is an upper limit. The model assuming IVR is close to 
case 2. A comparison of our model band fluxes (with IVR) with the Monte-Carlo simulations of \citet{Joblin2008}, in the case of the coronene cation (C$_{24}$H$_{12}^+$),
shows good general agreement, especially for the low-energy modes. We conclude that the IVR hypothesis provides an upper limit to the PAH vibrational emission at 
$\lambda>3$ mm and that the conclusions of the former section hold.

\section{Angular momentum distribution}
\label{j_dist}

Building the angular momentum distribution of interstellar PAHs is driven by photon exchanges and 
gas-grain interactions (DL98). In the DIM, pervaded by the ISRF, and for a PAH bearing 50 carbon atoms, the mean 
time between absorptions of visible-UV photons ($\sim$ 0.2 yr) is comparable to the mean time between emissions of rotational 
photons, as well as to the mean time between PAH-hydrogen collisions (for a gas density of 100 cm$^{-3}$ and a temperature of 100 K). 
Photons absorbed in the visible-UV have a weak effect on the total angular momentum; indeed, each photon exchanged carries a unit angular 
momentum, and the numerous\footnote{Energy conservation implies that 40 IR photons are emitted 
after each absorption.} IR photons emitted overwhelm the influence of the photon absorbed. To estimate the angular momentum 
distribution, we take the following processes into account:
\begin{itemize}
\item[-] IR rovibrational photon emission
\item[-] purely rotational photon emission
\item[-] H$_{2}$ formation
\item[-] collisions with gas (neutral atoms and ions)
\item[-] plasma drag
\item[-] photoelectric effect
\end{itemize}
All these processes lead to a change in the angular momentum of the molecule, some of them excite the rotation, and others damp it. 
For the radiative processes we adopt a quantum approach where each rotational state is treated individually 
and transition rules applied.
As discussed by \citet{Rouan97}, the building of the angular momentum distribution, $n(J)$, can be considered as a stationary random 
walk in a potential well with a minimum for $J = J_{0}$, for which the rate of $J$-change is zero:
\begin{equation}
\label{sum}
\left( \sum_{i} \left( \tau^{-1} \Delta J \right)_{i} \right)_{J_{0}} = 0 
\end{equation}
with $\Delta J$ the change of $J$ produced by the event number $i$, and $\tau$ the mean time between two events $i$. As in \citet{Rouan92}, we assume an 
efficient intramolecular vibration to rotation energy transfer (IVRET). We take $n(J)$ to be the same for all vibrational levels, equal to a Maxwell distribution 
$n(J)=n_0\,J^2\,\exp(-J^2/J_0^2)$, where $n_0$ is a normalization factor. Indeed, \citet{Mulas1998} and \citet{Ali2009} show that this form of $n(J)$ is 
a good approximation\footnote{With this form of $n(J)$ it is possible to define a rotational temperature (see Appendix \ref{temperature_rotationnelle}).}.  
In the following, we establish the rate of $J$-change due to the rovibrational emission of isolated PAHs modelled in Sect. \ref{p_de_e}. We also present 
the rate of purely rotational photon emission, while the contribution of the gas-grain interactions is described in Appendix \ref{gas_interaction}.

\subsection{Rovibrational transitions}
\label{rovib}

The rovibrational IR emission can be both an exciting and a damping process for the rotation of a PAH. Assuming that the 
interstellar PAHs are symmetric top molecules, the selection rules for the emission of an IR rovibrational photon are: $\Delta J = 0, 
\pm 1$, and $\Delta K = 0, \pm 1$. Transitions corresponding to $\Delta J = +1,-1$, and 0 are called the $P$, $Q$, and $R$-bands, respectively. 
As seen in Sect. \ref{pah_prop}, two types of transitions are possible: parallel ones 
(out-of-plane vibrational motion) with $\Delta K = 0$ and perpendicular ones (in-plane vibrational motion) with $\Delta K = \pm 1$. 
The type of each transition is given in Tables \ref{mid-IR} and \ref{far_IR}. For a rovibrational transition, $(v, J, K) \rightarrow (v-1, 
J+\Delta J, K+\Delta K)$, the rate is proportional to the spontaneous emission coefficient and to the probability for the grain to lose 
the corresponding transition energy. The spontaneous emission rate is proportional to the $A_{KJ}$ factors, which represent the angular part 
of the transition probability. Formulae for these factors are given in Appendix \ref{AKJ}. Expressed in terms of cross-section, the transition 
rate is (with $\nu_{i}$ and $\Delta\nu_{i}$ in cm$^{-1}$):
\begin{equation}
\label{IR_proba}
W^{\pm /0}_{i} = 8\pi c \; (\sigma_{i}\Delta\nu_{i}) \; \sum_{K=0}^{J} \; (\nu_{i}^{\pm /0})^{2} \; A_{KJ}^{\pm /0} \; P(h\nu_i^{\pm /0})
\end{equation}
where the $W^{\pm /0}_{i}$ give the transition rate for $\Delta J = 0, \pm 1$, and for the vibrational mode $i$ at a frequency $\nu_{i}^{\pm /0}$. This 
frequency depends on $\nu_{i0}$ (the frequency of the vibrational transition for $J=0$) and on $J$ and $K$ (see Appendix \ref{AKJ}). Since the $J$- 
and $K$-terms, in $\nu_{i}^{\pm /0}$, are always much smaller than $\nu_{i0}$, we take $P(h\nu_i^{\pm /0})\simeq P(h\nu_{i0})$. We note 
that the in-plane transitions provide higher (by a factor $\sim$ 2) rates than the out-of-plane ones, as a consequence of the $A_{KJ}$ sum values. The 
total transition rates are obtained by summing over all bands: $W^{\pm /0} = \sum_{i} \; W^{\pm /0}_{i}$. Finally, the rate of change in $J$ due to 
rovibrational transitions is $(\tau^{-1} \Delta J)_{IR} = W^{+}-W^{-}$. Rotational excitation dominates at low $J$, whereas damping 
is the dominant process for high $J$-values. For vibrational modes at higher frequency, $W^{+}-W^{-}$ is smaller
because the rotational terms are smaller in $\nu_i$ (see App. \ref{AKJ}):  $(\tau^{-1} \Delta J)_{IR}$ is therefore dominated by the contribution of
vibrational modes at low-frequencies.

\subsection{Rotational emission}
\label{rot}

We consider here the spontaneous emission of purely rotational photons. The selection rules for such transitions are $\Delta J = -1$ 
and $\Delta K = 0$ \citep{Townes}. No change in $K$ occurs because, for rotational transitions, the dipole moment of a symmetric top 
molecule necessarily lies along its symmetry axis. The transition rate is then simply related to the spontaneous emission coefficient 
$A_{J, J-1}$: 
\begin{eqnarray}
\label{einstein}
(\tau^{-1} \Delta J)_{rot} &=& - A_{J,J-1} \\
A_{J,J-1} &=& \frac{512\pi^{4}}{3h^4c^3} B^3\mu^2 J^{3} \frac{(2J+1)^{2}-(J+2)}{3(2J+1)^2} \; \; \ \rm{s}^{-1}
\end{eqnarray}
where $\mu$ is the electric dipole moment of the molecule and the factor $A_{KJ-}$ has been used (Appendix \ref{AKJ}). In the high-$J$ limit, 
$h\nu\times A_{J,J-1}$ tends to the classical expression of Larmor. Finally, the damping rate due to rotational emission is\footnote{In the limit $J \gg 1$, we estimated that the stimulated 
emission and absorption of CMB photons represent less than 20\% of the spontaneous emission rate.}
\begin{eqnarray}
\label{AJJ-1}
(\tau^{-1} \Delta J)_{rot} = &-& 1.8 \times 10^{-14} \left(\frac{N_{C}}{50}\right)^{-6}\left(\frac{\mu}{1\,\rm{D}}\right)^{2} \nonumber \\
						     &\times& J^{3} \frac{(2J+1)^{2}-(J+2)}{(2J+1)^2} \; \; \ \rm{s}^{-1} \; .
\end{eqnarray}

\subsection{Equilibrium angular momentum $J_0$}
\label{paragraph_contrib}

\begin{table}[t]
\centering
\begin{minipage}{0.5\textwidth}
\centering
\caption{Physical parameters for the typical interstellar phases considered, $n_{H}$ the hydrogen density, $T$ the gas 
temperature, $n_{e}$ the electrons density, $n_{H^{+}}$ the proton density, and $n_{C^{+}}$ the density of atomic carbon ions.}
\label{environments}
\centering
\begin{tabular}{cccccc}
\hline
\hline
  & MC \footnote{Molecular cloud.}  & CNM\footnote{Cold neutral medium.} & WNM\footnote{Warm neutral medium.}  & WIM\footnote{Warm ionized medium.}  & Orion Bar \\
\hline
$G_{0}$                   & $0.01$    & $1$                  & $1$                & $1$        & $14\,000$          \\
$n_{H} \; (\rm{cm}^{-3})$ & $300$     & $30$                 & $0.4$              & 0.1        & $10^{4}$           \\
$T \; (\rm{K})$     & $20$      & $100$                & $6000$             & 8000       & $400$              \\
$n_{e} \; (\rm{cm}^{-3})$ & $0.03$    & $0.045$              & $0.04$             & 0.1        & $3$                \\
$n_{H^{+}}/n_{H}$         & $0$       & $1.2 \times 10^{-3}$ & $0.1$              & 0.99        & $10^{-4}$          \\
$n_{C^{+}}/n_{H}$         & $10^{-4}$ & $1.3 \times 10^{-4}$   & $1.3 \times 10^{-4}$ & 10$^{-3}$  & $2 \times 10^{-4}$ \\
\hline
\end{tabular}
\end{minipage}
\end{table}

\begin{figure}[!t]
\centering
\resizebox{\hsize}{!}{\includegraphics[angle=90]{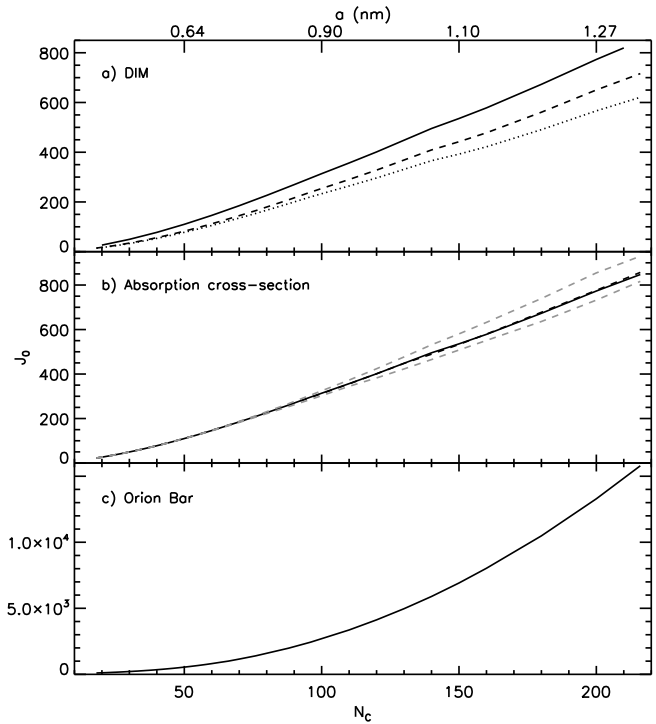}}
\caption{$J_{0}$ versus $N_C$ for PAH cations. In a) we show the case of the diffuse medium. Solid line 
shows the CNM, dashed line the WNM and dotted line the WIM. In b) we illustrate the influence of the absorption cross-section in the CNM: 
cations (solid line), neutrals (dashed line) and extreme behaviours (grey lines) for the first vibrational mode versus $N_C$ 
(see Fig. \ref{E_weight}) are shown. In c), we present the case of the Orion Bar.}
\label{Jo_Nc}
\end{figure}

\begin{figure*}[!htbp]
\centerline{
\begin{tabular}{cc}
\includegraphics[width=8.5cm]{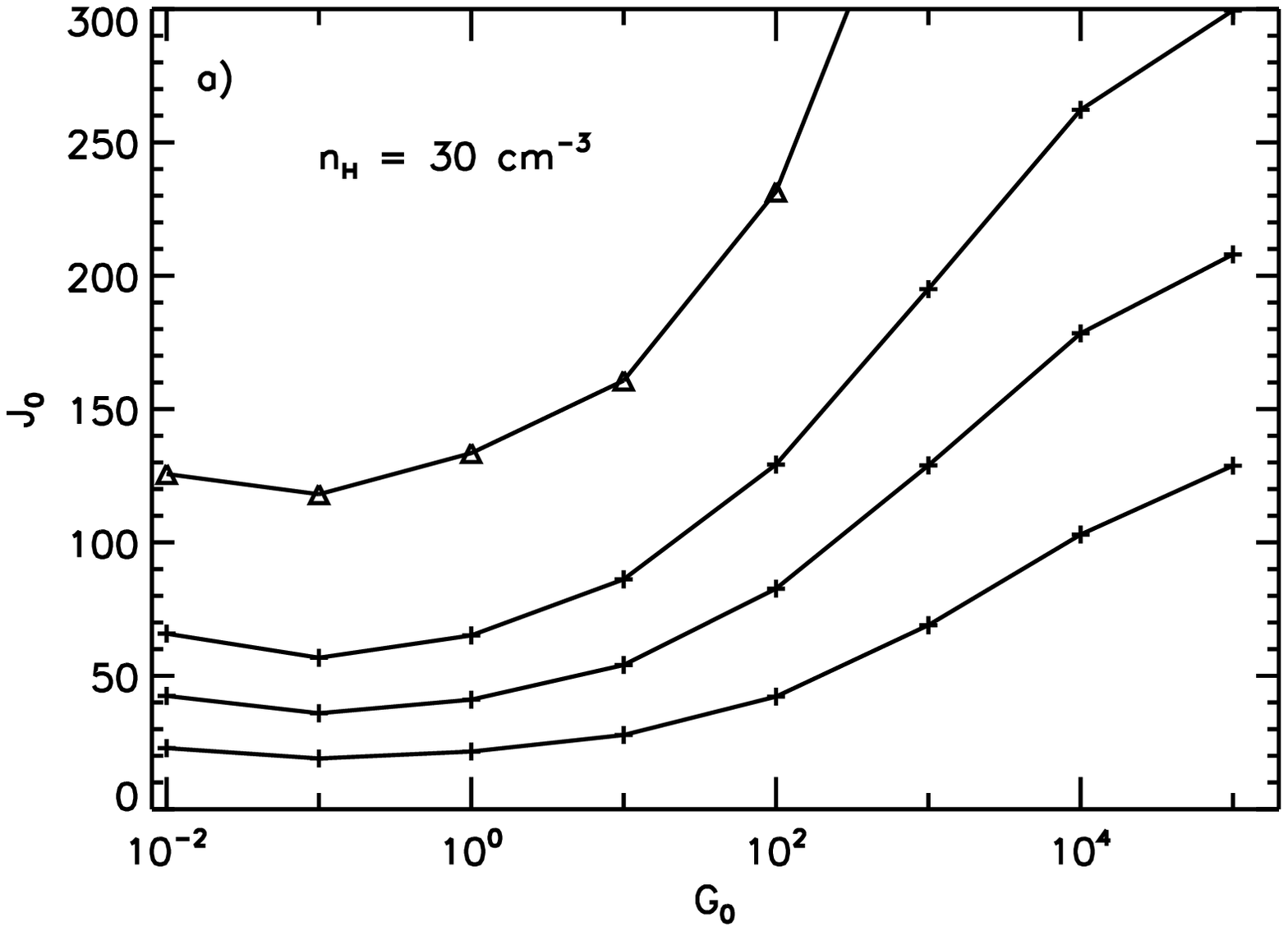} & \includegraphics[width=8.5cm]{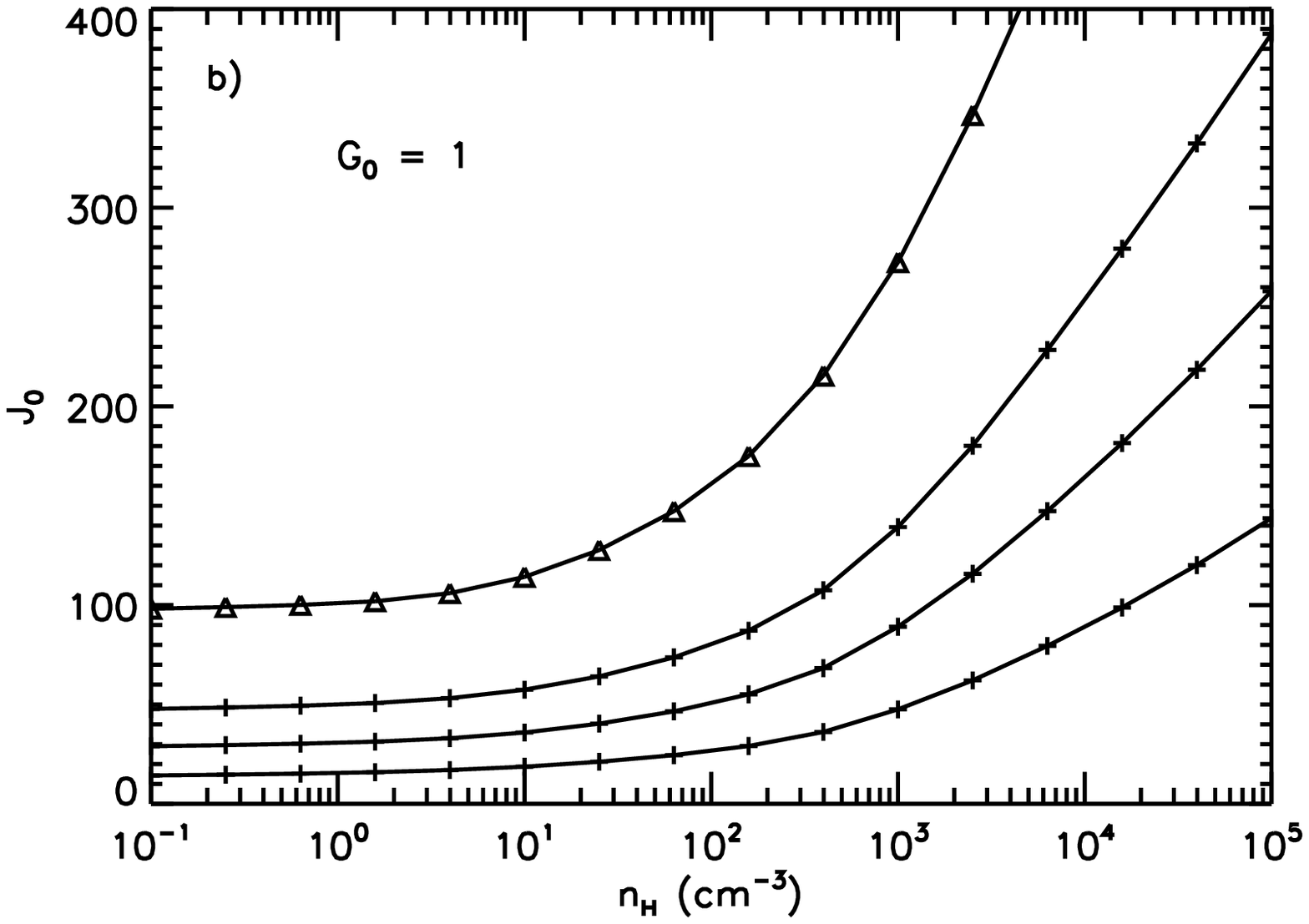}
\end{tabular}}
\caption{{\em a)} Effect of varying $G_0$ on $J_0$ for PAH cations with $N_C$ = 20, 30, 40, and 100 (from 
bottom to top). The values for $N_C = 100$ (triangles) have been divided by 2. The gas density is 30 cm$^{-3}$ 
and the other physical parameters for the gas are determined using CLOUDY \citep{CLOUDY}. {\em b)} Same as a) 
but for varying $n_H$ at $G_0 = 1$. In both cases the radiation field is a blackbody with $T_{eff} = 22\,000$ K
and the gas parameters have been obtained with CLOUDY at thermal equilibrium (see text).}
\label{chi} 
\end{figure*}

\noindent We estimate damping and excitation rotational rates for the interstellar phases described in Table \ref{environments}. When solving Eq. \ref{sum}, we 
obtain the equilibrium angular momentum $J_0$. Figure \ref{Jo_Nc} shows $J_0$ for the DIM and the Orion Bar. In all cases,  $J_0$ increases with 
$N_C$ because there are more steps in the random walk; indeed, the cross-section and absorption rate scale with $N_C$ (see Fig. \ref{sigma}), so 
more IR photons are emitted for larger $N_C$. Similarly, $J_0$ reaches higher values when the radiation field intensity or the gas density increases
(see Fig. \ref{chi}).

\noindent In the DIM, the excitation comes from gas-grain collisions, whereas the damping is dominated by the emission of rotational 
photons for $N_C < 140$ and by IR photon emission for larger PAHs ($(\tau^{-1}\Delta J)_{rot} \propto N_C^{-5}$, whereas $(\tau^{-1}\Delta J)_{IR} \propto N_C$). 
We note similar $J_0$-values for the CNM, WNM and WIM (Fig. \ref{Jo_Nc}a) in spite of the large differences in gas density and temperature
(see Table \ref{environments}). In the Orion Bar, because of the intense radiation field, the excitation is driven by the photoelectric effect, whereas the 
damping is dominated by the emission of IR photons. For the case of MCs (not shown), $J_0$ also rises 
and reaches high values ($J_0=1300$ for $N_C=200$). Indeed, the damping by IR photon emission is no longer efficient ($G_0\sim 0.01$), while the gas-grain 
rates are strong because of higher density. We show in Fig. \ref{Jo_Nc}b the influence of different 
choices of the vibrational mode spectrum (or cross-section) on $J_0$: changes are below 20\% and affect only the large sizes ($N_C>100$), which make a minor contribution
to the rotational emissivity (see Sect. \ref{spin_par}). We found comparable variations of $J_0$ for the vibrational relaxation (IVR) cases 1 and 2, discussed in 
Sect. \ref{decouplage}.

\noindent The processes contributing to the rotation of PAHs depend on the intensity of the radiation field ($G_0$) and on the gas 
density ($n_H$). We now examine the influence of these parameters on $J_0$. Other important quantities (gas temperature; electron, proton,
and C$^+$ abundances) are obtained at thermal equilibrium with CLOUDY \citep{CLOUDY}\footnote{We assumed 130 (320) ppm of carbon (oxygen) to be in the gas phase. 
Parameters were taken from the optically thin zone of isochoric simulations with CLOUDY.}. 
To study the influence of $G_0$, we took the radiation field to be a blackbody of effective temperature 22,000 K and scaled it in order to vary $G_0$ 
between $10^{-2}$ and $10^5$. The shape of the radiation field is then always the same\footnote{We also varied the shape of the blackbody keeping $G_0$ constant (as well as 
$n_H$). Similar $J_0$ values were found for $T_{eff}$ between 10$^4$ and $5\times10^4$ K.}. Figure \ref{chi}a shows $J_0$ as a function of $G_0$ for different PAH sizes. 
As discussed before, $J_0$ increases with $G_0$ because the excitation rate from IR emission (which scales with $G_0$) is dominant. This rise becomes steeper for larger 
PAHs because the frequency of the first vibrational mode decreases (as $N_C^{-1}$) and requires lower values of $G_0$ to be excited (the maximum of $P(E)$ moves to higher 
energies as $G_0$ increases, see Fig. \ref{P_E}c). However, we note that $J_0$ is affected little by variations of $G_0$ over the range 0.01-100.
The result of varying $n_H$ is shown in Fig. \ref{chi}b. In all cases, the incident radiation field is a blackbody with $T_{eff}$ = 22,000 K and $G_0 =1$.
The gas-grain processes become dominant for $n_H>30$ cm$^{-3}$ and $N_C > 50$. At lower gas densities, radiative processes prevail (excitation by IR emission and damping by 
rotational emission), and $J_0$ does not change. We therefore expect the rotational excitation of small PAHs to hardly be variable in the DIM as found by \citet{Davies2006}.

\subsection{Modelling the rotational excitation by IR emission}

\begin{figure}[!t]
\centering
\resizebox{\hsize}{!}{\includegraphics{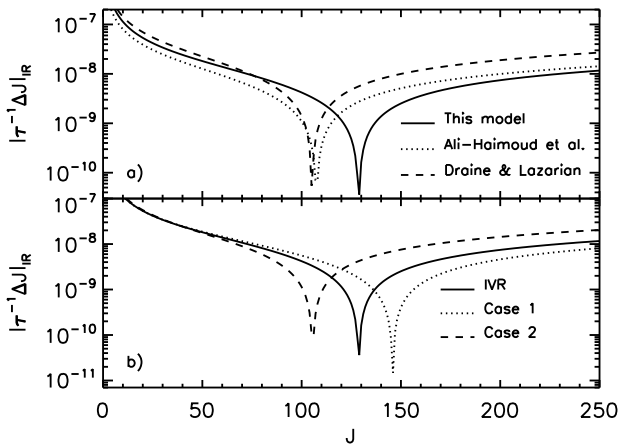}}
\caption{Rate of $J$-change due to rovibrational emission for a PAH with $N_C = 24$ and heated by the ISRF. {\em a)} Comparison with former models: our rate (solid line), the rate of DL98 
corrected as in \citet{Ali2009} (dashed line) and the rate of \citet{Ali2009} (dotted line). {\em b)} Effect of IVR breakdown, case 1 and 2 (see Sect. \ref{decouplage}).}
\label{comparaison_taux}
\end{figure}

\noindent Following the absorption of a stellar visible-UV photon, the rovibrational cascade of PAHs is a complex process that involves many states. Molecular, state-to-state 
models require a detailed database, use Monte-Carlo simulations (see Mulas et al. 2006 for references), and so far do not include the gas-grain interactions that are important. 
Given our incomplete knowledge of interstellar PAHs and the fact that their rotational emission may so far be seen in broadband data, former models of spinning dust 
(DL98 and Ali-Ha{\"i}moud et al. 2009) made simplifying assumptions to describe radiative processes and performed a classical treatment of the gas-grain interactions
\footnote{These models can also be applied to describe the rotational emission of other grain types.}. We assess here the impact of these assumptions on the rotational emissivity of PAHs.

\noindent First, the internal energy distribution of PAHs was derived in the thermal approximation. As discussed in Sect. \ref{P_E}, this is a questionable assumption for describing 
the long-wavelength emission of PAHs and the change in angular momentum it induces. Moreover, the excitation rate by IR emission (the recoil due to emission of individual photons), which 
is a purely quantum effect, has been described as a random walk of the angular momentum starting with a non-rotating grain\footnote{Conversely, the damping part is correctly described by a 
classical model when $J>>1$ (Ali-Ha{\"i}moud et al. 2009).}. Finally, an efficient vibrational redistribution (IVR, see Sect. \ref{decouplage}) was assumed throughout the energy cascade 
following a photon absorption. In this work, we improve on the first aspect by deriving the internal energy distribution of isolated PAHs using a microcanonical formalism and including the 
rotational density of states. Next, we follow a quantum approach to treat the rovibrational emission where the recoil due to photon emission and selection rules are naturally included. In Fig. 
\ref{comparaison_taux}a, we compare the absolute value of our rate of angular momentum change by rovibrational IR emission  $|(\tau^{-1}\Delta J)_{IR}| = |W^+ - W^-|$ (see Sect. 
\ref{rovib}) to former works. For a PAH containing 24 carbon atoms, all rates decrease with the angular momentum and cross zero for $J$ between 100 and 150 (singular points in our 
logarithmic representation). The main difference with previous models is that the excitation-to-damping transition (the zero value) occurs at higher $J$-values in our case. This discrepancy 
diminishes with increasing sizes, and identical IR rates are found for species with more than 100 C atoms.

\noindent We show in Fig. \ref{comparaison_taux}b the effect of relaxation schemes other than IVR. Between extreme cases 1 and 2 (see Sect. \ref{decouplage}), the damping IR 
rate is multiplied by a factor 2, which leads to a balance at a lower $J$-value. Variations in $(\tau^{-1}\Delta J)_{IR}$ are in fact quite similar to those induced by the other approximations 
discussed above. All these variations affect the rotational emissivity of PAHs, around 30 GHz, by at most 15\% (in the CNM and including the gas-grain interactions)\footnote{A correction 
of the same order was found by \citet{Ali2009}, who used the Fokker-Planck equation to derive $n(J)$ instead of a Maxwell distribution.}. We conclude that, in spite of 
the assumptions made, former spinning dust models (which are fast computationally) provide a sufficiently accurate rotational emissivity.

\begin{figure*}[!htbp]
\centerline{
\begin{tabular}{cc}
\includegraphics[width=8.5cm]{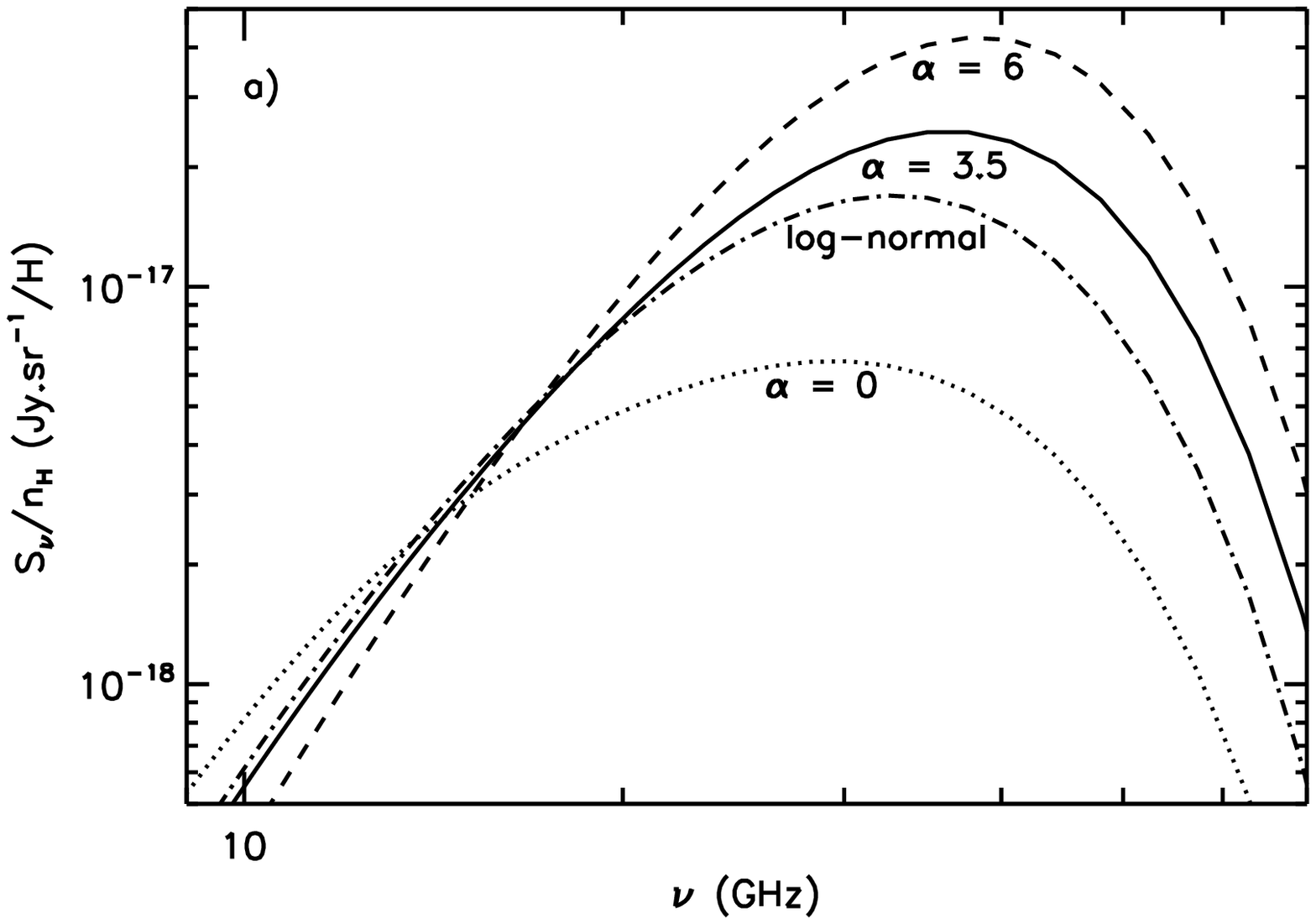} & \includegraphics[width=8.5cm]{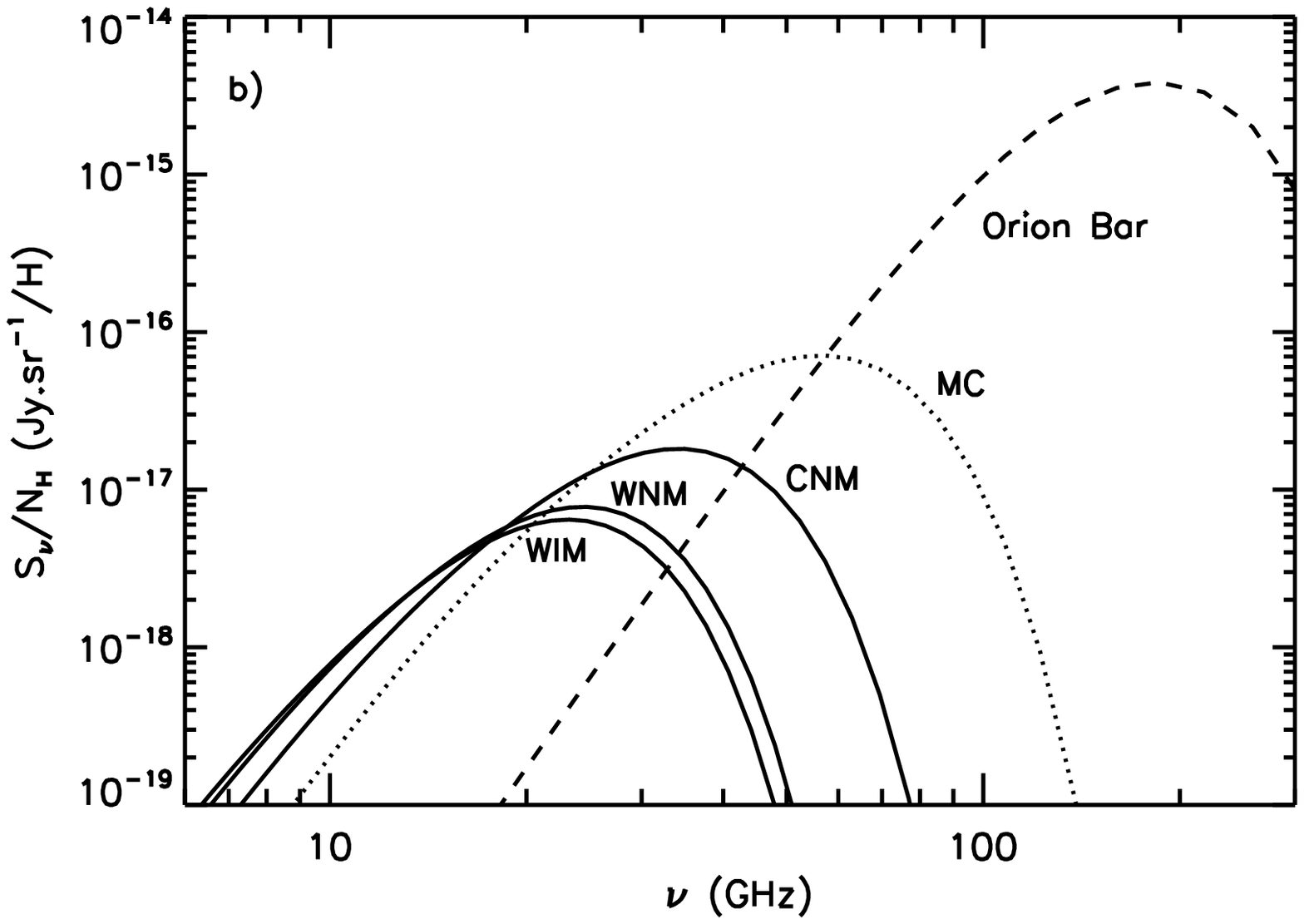}
\end{tabular}}
\caption{Rotational emissivity of PAH cations. We take 430 ppm of carbon in PAHs, $N_{C}=24-216$ 
and assume $m = 0.4$ Debye. {\em Panel a):} effect of changing the fraction of large PAHs (decreasing with
$\alpha=0$ to 6). The case of a log-normal distribution centered around $N_C=44$ and width 0.4 is also shown. {\em Panel b):} rotational emission spectrum 
in different environments with a power law size distribution and $\alpha=3.5$.}
\label{radio}
\end{figure*}

\section{Spinning dust emission}
\label{spin_par}

The power emitted by a PAH containing $N_C$ carbon atoms in a rotational transition from state $J$ to $J-1$ is equal to
\begin{equation}
P(J) = A_{J,J-1} \times 2BJ \; ,
\end{equation}
when taking the spontaneous emission rate $A_{J,J-1}$ from Eq. \ref{einstein} and with the transition energy $h\nu=2BJ$ ($\Delta K=0$ 
for rotational transitions). With $B\propto N_C^{-2}$ and $\mu^2\propto N_C$, we find that $P(J)\propto N_C^{-7}$: small PAHs will therefore have a dominant contribution 
to the rotational emission. For instance, the peak value of the rotational emissivity of a PAH with $N_C=96$ is 10 times lower than for a species with $N_C=24$ (see also 
Ali-Ha{\"i}moud et al. 2009). With $n_{s}$, the size distribution of PAHs (the number of PAHs of a given size $a$ or $N_C$ per proton), and the angular momentum distribution 
$n(N_{C},J)$, we get
\begin{eqnarray}
\label{eq_radio}
S_{\nu} = \frac{N_{H}}{4\pi} \int_{N_{min}}^{N_{max}} A_{J \rightarrow J-1} \; n(N_{C},J) \; \frac{2BJ}{2Bc} \; n_{s}(N_{C}) \; dN_{C} \; .
\end{eqnarray}
Since we are interested in a broadband spectrum, we take the rotational bandwidth to be $2Bc$. As discussed before, we assume that $n(N_C,J)$ 
is described correctly by a Maxwell distribution. Given the relationship $J_0(N_C)$ (Fig. \ref{Jo_Nc}) and the size distribution, the emission of 
spinning PAHs can be calculated from Eq. \ref{eq_radio}. It scales with $m^2\times\, S_{PAH}$. We illustrate the relative significance of 
small species in Fig. \ref{radio}a by changing the index of the power law size distribution, $n_s(a)\sim a^{-\alpha}$. 
Defining small PAHs by $N_C\lesssim 100$, they represent 50\% (97\%) of the total 
abundance for $\alpha=0$ (6). As expected, the rotational emissivity for $\alpha=6$ is highest. It is also blueshifted by 5 GHz
and broadened by about the same amount with respect to the $\alpha=0$ case.
Figure \ref{radio}b shows the case of different interstellar environments and illustrate the influence of 
the gas-grain processes, which become strong in the CNM (while subdominant in the WIM/WNM).

\noindent We emphasize here that the level, peak position and width of the rotational emission spectrum depend on (a) the fraction of small PAHs ($N_C\leq 100$), 
(b) $m$ the scaling factor for the dipole moment, 
and (c) the physical parameters of the gas along the line of sight (temperature, density, and ionization). Microwave observations alone cannot constrain all these unknowns. 
A quantitative description of the spinning dust emission will require near-IR data (to constrain the fraction of small PAHs) and radio maps (21 cm and continuum) to 
derive the physical state of the gas.

\section{The case of the molecular cloud G159.6-18.5}
\label{observations}

\begin{figure}[!t]
\centering
\resizebox{\hsize}{!}{\includegraphics{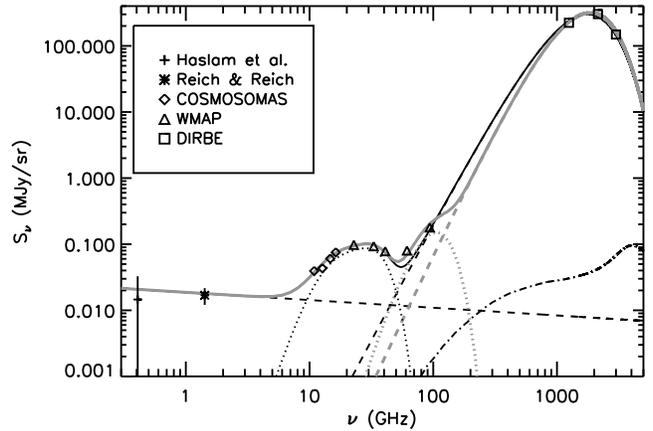}}
\caption{Comparison to observations (symbols) of the Perseus molecular cloud G159.6-18.5. The dashed lines show the free-free (low frequency) and big grains (high-frequency) 
emissions. The PAH rovibrational emission is represented by the dot-dashed line. Dotted lines show the PAH rotational emission. The black model is model A, and the grey model is 
model B. See text for details.}
\label{cosmosomas}
\end{figure}

\noindent We present here a comparison of our model results with observations of the anomalous microwave foreground and IR emission. We show how these data 
can be explained with a coherent description of the IR and rotational emission of PAHs, and also discuss how the size distribution of PAHs can be constrained.

\noindent The COSMOSOMAS experiment has delivered maps of large sky fractions with an angular resolution of 1 degree in four frequency bands: 
10.9, 12.7, 14.7, and 16.3 GHz. In this survey, \citet{Watson} studied the region G159.6-18.5 located in the Perseus molecular complex. Figure \ref{cosmosomas} 
shows the spectrum of this region from 3 to to 4000 GHz (COSMOSOMAS, WMAP, and DIRBE data). From mid to far-IR, the Perseus molecular complex 
is dominated by a molecular ring (G159.6-18.5) surrounding an HII region. Unfortunately, the COSMOSOMAS resolution does not allow these two phases to be separated. 
The molecular ring is centred on the star HD278942 \citep{Andersson2000, Ridge2006}. Its stellar wind is responsible for the HII region 
dug into the parent molecular cloud. To describe the emission spectrum of G159-18.5, we assume it includes an MC component for the ring and a WIM component for 
the HII region. We discuss below the physical parameters of both phases.

\noindent  According to \citet{Ridge2006}, HD278942 is a B0 V star: its radiation field is modelled by a blackbody with $T_{eff} \sim$ 30,000 K, scaled to get 
$G_0 \sim 1.6$ at the radius of the ring ($G_0$ is obtained from IRAS data). The radius of the HII region is approximately 2.75 pc \citep{Andersson2000} 
and implies a WIM density $n_H\sim 1$ cm$^{-3}$ \citep{Ridge2006}.  We obtained the WIM temperature from CLOUDY: $7\,500$ K. \citet{Andersson2000} 
estimated the density and temperature of the gas in the molecular ring to be $n_H \sim 560$ cm$^{-3}$ and $T = (20\pm10)$ K. For the WIM and MC phases, we derived the 
abundances of electrons, protons and C$^+$, and the gas temperature with CLOUDY. We took the total column density (MC+WIM) to be $N_H = 1.3\times 10^{22}$ cm$^{-2}$ 
\citep{Watson}. From 12 $\mu$m IRAS data and our PAH emission model, we estimated the total PAH abundance to represent $(240\pm 40)$ ppm of carbon. 

\noindent Following \citet{Watson}, we took the thermal dust emission to be a grey body at 19 K with an emissivity index $\beta=1.55$. Our best-fit parameters for the spinning PAH
contribution (assuming a size distribution $n_s(a)\sim a^{-3.5}$) are then ({\em model A}) $N_C=50-216$ and $N_H=4.6 \times 10^{21}$ cm$^{-2}$ for the MC, $N_C=24-216$ 
and $N_H=8.1\times 10^{21}$ cm$^{-2}$ for the WIM. In both media, we took $m=0.6$ D (Eq. \ref{equation_mu}). Larger PAHs are required in the MC, possibly as a 
consequence of grain-grain coagulation. However, the results of \citet{Dupac2003} indicate that emission of dust around 20 K requires a higher emissivity index, close to 2. 
When adopting $\beta=2$ and a dust temperature of 18 K ({\em model B}), we find an equally good fit of this region in the far-IR (Fig. \ref{cosmosomas}). In this case, the flux around 
100 GHz is not explained. We speculate here that it may come from an additional and less abundant population of small PAHs in the MC component with: $N_C=24-50$, $m=0.1$ D and 
$N_H = 1.25 \times 10^{21}$ cm$^{-2}$ (for the larger PAHs in the MC phase, $N_H = 4.25\times 10^{21}$ cm$^{-2}$). The corresponding spinning emission is shown in Fig. \ref{cosmosomas} 
({\it model B}). The PAH size distribution would thus be bimodal, as already suggested by \citet{LePage}. As can be seen in Fig. \ref{cosmosomas}, both models provide a good fit to the data. 
Discrimiting tests of these scenarios will be soon possible with the Planck polarized data. Small grains or PAHs are expected to be poorly aligned 
(Lazarian \& Draine 2000, Martin 2007), whereas big grains are fairly well-aligned with a polarization fraction of 5 to 10\%.

\section{Summary}

\noindent The Planck and Herschel data will soon reveal the emission of interstellar dust at long wavelengths. 
Thanks to their small size, interstellar PAHs spend most of their time at low internal energy and can spin at frequencies of a few tens of GHz. The emission of PAHs
is therefore expected to make a significant contribution at long wavelength. Recent observations have shown 
the existence of a 10-100 GHz emission component (the anomalous foreground), related to the smallest dust grains. As suggested by DL98, the anomalous foreground
may trace the emission of spinning PAHs.

\noindent In this work, we built the first model that coherently describes the emission of interstellar PAHs from the near-IR to the centimetric range and focused on the 
long wavelength part of this emission. To do so, we derived the internal energy of isolated PAHs down to low energies. We included low-frequency vibrational bands ($\lambda > 20\;\mu$m), 
through which PAHs cool at intermediate to low energies. They are important for their rotational emission. In the cooling cascade that follows the absorption of a stellar 
photon by a PAH, we treated rovibrational transitions in a quantum approach and examined the possibility that the hypothesis of vibrational redistribution (IVR) is not 
always fulfilled. Purely quantum effects (recoil due to photon emission; transitions that do not change the angular momentum or $Q$-bands) are thus naturally included in our 
description of the rovibrational cascade. We obtained the rotational emission of PAHs from a random-walk formalism, including all processes participating in excitation 
and damping, namely, rovibrational and pure rotational transitions and gas-grain interactions.

\noindent We have shown that the rovibrational emission of PAHs above $\sim$ 3 mm does not depend on the intensity of the radiation field (represented by $G_0$),
unlike the mid-IR part of the spectrum that scales linearly with it. In the diffuse interstellar medium, PAHs may contribute up to 10\% of the dust emission around 
100 GHz. We also found the rotational emissivity of PAHs is dominated by small species (bearing less than 100 C atoms) and is hardly sensitive to $G_0$ over the range 0.1-100. 
Using plausible PAH properties, our model can explain both the IR and microwave emissions of a molecular cloud in the Perseus arm where the anomalous foreground is conspicuous.
The level, peak position and width of the rotational emission spectrum depend on the fraction of small PAHs, the dipole moment distribution ($m$-factor), and the physical parameters of 
the gas phases present along the line of sight. A quantitative description of the emission of spinning PAHs will therefore involve observations at IR and radio wavelengths (21 cm and 
continuum). Comparing the rotational excitation rate obtained from our quantum treatment of the rovibrational cascade to former works, we showed that the classical approximation used so far  
has little effect on the rotational emissivity (the peak value varies by at most 15\%). Similarly, departures from the IVR hypothesis lead to similar emissivity changes. We therefore 
conclude that a classical description of rovibrational transitions and the IVR hypothesis are good approximations for describing the rotational emission of PAHs.

\noindent Our results on the influence of the radiation field intensity led to a specific prediction that can be tested observationally. If the anomalous microwave foreground comes from 
spinning PAHs, it is expected to be correlated with the dust emission in the 12 $\mu$m-IRAS band, mostly carried by PAHs. In regions where $G_0$ varies significantly, this correlation 
should improve when the 12 $\mu$m flux is divided by $G_0$, indeed, the IR emission of PAHs scales with $G_0$, whereas their rotational emission is independent of $G_0$.
This prediction was tested in a companion paper with WMAP and IRAS data \citep{Ysard2009b}.

\acknowledgements{We thank an anonymous referee whose comments helped us to significantly improve the content of this paper. We gratefully acknowledge stimulating discussions with 
B. Draine, C. Joblin, E. Dartois, T. Pino, and O. Pirali. We are grateful to M. Compi\`egne for his help in the dust SED modelling.}

\bibliographystyle{aa} 
\bibliography{biblio}

\begin{appendix}

\section{Mid-IR absorption cross-section}

\label{appendix_sigma}

In Table \ref{mid-IR} we give the parameters defining the mid-IR vibrational bands considered in this work. In all cases we assume their profile to have a Drude shape. 
The bands at 3.3, 6.2, 7.7, 8.6, 11.3, and 12.7~$\mu$m are defined as follows. For the PAH cations, we start from the integrated cross-sections, $\sigma\Delta\nu$, of 
\citet{Pech2002} that have been derived from laboratory data. The corresponding band profiles, however, do not provide a detailed match of observations. We therefore 
use band positions and widths as deduced from fits of ISO-SWS spectra of a number of interstellar regions \citep{Verstraete01}. As indicated by these observations 
and others \citep{Peeters2002}, we include a broad band at 6.9 $\mu$m and split the 7.7~$\mu$m into three sub-bands at 7.5, 7.6, and 7.8~$\mu$m, where we use the 
observed $\sigma \Delta\nu$ of each sub-band as weights in defining their integrated cross-sections. In addition, we introduce a band at 8.3 $\mu$m to fill the gap 
between the 7.8 and 8.6 $\mu$m bands, and multiply the 8.6~$\mu$m band by a factor 3 to match observations. For neutral PAHs, we use the laboratory 
integrated cross-section of \citet{Joblin1995}, and assume the same band profiles as for the cations. Furthermore, spectroscopic data (ISO-SWS, Spitzer
-IRS and UKIRT) reveal other bands at 5.25 and 5.75 $\mu$m, which have been ascribed to combinations of PAH vibrational modes involving the 11.3~
$\mu$m band and IR-forbidden modes at 9.8 and 11.7~$\mu$m, respectively \citep{Roche1996, Tripathi}. For the 5.25~$\mu$m, we use the width and intensity ratio to 
the 11.3~$\mu$m band given in \citet{Roche1996}. The 5.75~$\mu$m band has been derived from the observed spectrum of the Orion Bar \citep{Verstraete01}. 
We also add the 17.1~$\mu$m band recently seen in Spitzer data \citep{Smith2004,Werner2004} and recognized as arising from PAHs 
\citep{Peeters2004,Smith07}.

\begin{table}[t]
\begin{minipage}{0.5\textwidth}
\centering
\caption{Mid-IR bands of interstellar PAHs adopted in this work for cations and neutrals.}
\label{mid-IR}
\centering
\begin{tabular}{cccccc}
\hline
\hline
$\lambda_{i}$  & $\nu_{i}$        & $\Delta \nu_{i}$ & $\sigma_{i}/N_H$              & $\sigma_{i}/N_H$               & Type\footnote{In-plane (ip) or out-of-plane (op) character of the bands (see Sect. \ref{pah_prop}).} \\
$\; (\mu$m$)$  & $(\rm{cm}^{-1})$ & $(\rm{cm}^{-1})$ & $(\rm{10^{-20}cm^{2}})$ & $(\rm{10^{-20}cm^{2}})$  &      \\
               &                  &                  &  cations                  & neutrals                   &      \\
\hline
3.3 & 3040 & 39 & 2.44 & 10.8 & ip \\
5.2 & 1905 & 23 & 0.58 & 0.58 & op \\
5.7 & 1754 & 60 & 0.49 & 0.49 & op \\
8.3 & 1205 & 63 & 1.74 & 1.74 & ip \\
8.6 & 1162 & 47 & 5.34 & 0.51 & ip \\
11.3 & 890 & 18 & 17.3 & 18.3 & op \\
12.0 & 830 & 30 & 3.17 & 3.17 & op \\
12.7 & 785 & 16 & 5.06 & 4.06 & op \\
\hline
\hline
$\lambda_{i}$  & $\nu_{i}$        & $\Delta \nu_{i}$ & $\sigma_{i}/N_C$              & $\sigma_{i}/N_C$              & Type \\
$\; (\mu$m$)$  & $(\rm{cm}^{-1})$ & $(\rm{cm}^{-1})$ & $(\rm{10^{-20}cm^{2}})$ & $(\rm{10^{-20}cm^{2}})$ &      \\
               &                  &                  &  cations                  & neutrals                  &      \\
\hline
6.2  & 1609 & 44  & 2.48 & 0.52 & ip \\
6.9  & 1450 & 300 & 0.40 & 0.40 & ip \\
7.5  & 1328 & 70  & 2.70 & 0.12 & ip \\
7.6  & 1315 & 25  & 1.38 & 0.06 & ip \\
7.8  & 1275 & 70  & 2.70 & 0.12 & ip \\
16.4 & 609  & 6   & 1.83 & 1.83 & ip \\
17.1 & 585  & 17  & 2.48 & 2.48 & ip \\
\hline
\end{tabular}
\end{minipage}
\end{table}

\section{Vibrational modes and density of states of interstellar PAHs}

\label{modes}

\subsection{Vibrational modes}

\begin{figure}[!t]
\resizebox{\hsize}{!}{\includegraphics[angle=90]{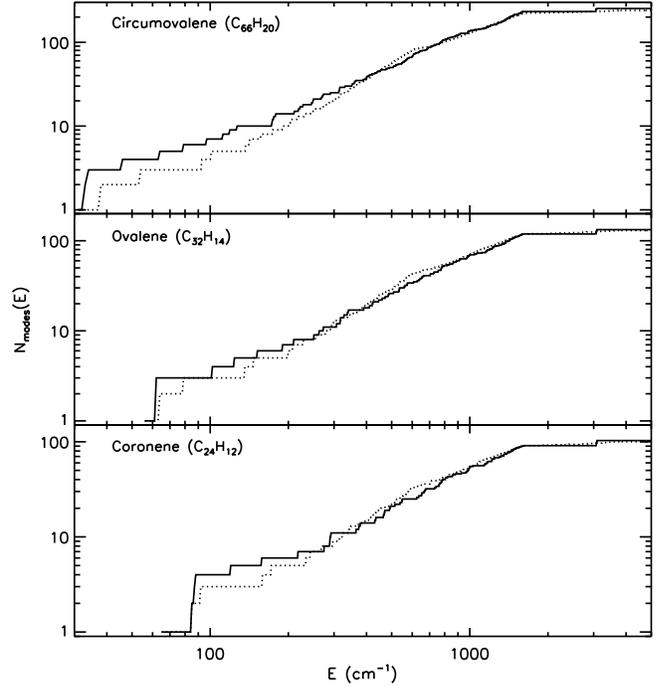}}
\caption{Cumulative distribution of vibrational modes for coronene ($C_{24}H_{12}$), ovalene ($C_{32}H_{14}$) and circumovalene 
($C_{66}H_{20}$), from the \citet{Malloci2007} database (solid lines) and with our Debye model (dotted lines).}
\label{normal_modes}
\end{figure}

Symmetric top ($D_{6h}$ symmetry) type PAHs, with $N_{C}$ carbon atoms and $N_{H} = \sqrt{6N_{C}}$ hydrogen atoms, have 
$3(N_{C}+N_{H}-2)$ vibrational modes that can be divided into the following types: $(N_{C}-2)$ out-of-plane ($op$) CC modes, 
$2(N_{C}-2)$ in-plane ($ip$) CC modes, $N_{H}$ out-of-plane CH bending modes, $N_{H}$ in-plane CH bending modes, and $N_{H}$ CH 
stretching ($st$) modes. Following \citet{Li}, we approximate the mode spectrum of each type of vibration with a two-dimensional 
Debye model of maximum energy $k\Theta$, where $\Theta$ is the Debye temperature. We derive the mode spectra from the following 
expressions:
\begin{enumerate}
\item[-] for the CC modes:
\begin{equation}
\hbar\omega_{i} = k\Theta_{t} \sqrt{\frac{i-\delta^{t}_{i}}{N_{t}}} \;\, {\rm for} \;\, i = 1, N_{t}
\end{equation}
\noindent where $t=op,ip$ is the type of mode and $N_{t}$ is the number of CC modes of a given type with $\Theta_{op} = 876$ K and 
$\Theta_{ip} = 2318$ K with:
\begin{eqnarray}
\delta^{op}_{i} &=& 3/2 \; {\rm for} \; \, i = 2, 3 \;\;\;\;\; {\rm and} \;\;\;\;  \delta^{ip}_{i} = 1 \; {\rm for}\; \, i = 2, 3 \nonumber \\
&=& 3/4 \; { \rm otherwise} \;\;\;\;\;\;\;\;\;\;\;\;\;\;\;\;\;\;\;\;\, = 1/2 \; \rm{otherwise} \nonumber \\
\end{eqnarray}
\item[-] for the CH modes:
\begin{equation}
\hbar\omega_{i} = k\Theta_{t} \sqrt{\frac{3i}{2N_{t}}} \;\, {\rm for} \;\, i = 1, N_{t}
\end{equation}
\noindent where $t=op,ip, st$, and $N_{t}$ is the number of CH modes of a given type with $\Theta_{op} = 1281$ K, $\Theta_{ip} = 1672$ K 
and $\Theta_{st} = 4375$ K.
\end{enumerate}
To this mode spectrum, we add the mode at lowest energy as described in Sect. \ref{abs_cross_sect}. Figure \ref{normal_modes} shows that 
our Debye mode spectra agree with that derived from the \citet{Malloci2007} database.

\subsection{Density of states and degeneracies}

We estimate the PAH density of harmonic vibrational states from direct counts, a mode-to-mode convolution method proposed by 
\citet{Beyer}. We start from the rotational density of states, classically given by $\rho_{rot} (E) = \sqrt{8E}/B^{3/2}$
with $E$ and $B$ in cm$^{-1}$ and for a symmetric top \citep{Baer}. We then obtain the rovibrational density of states $\rho(E)$ with 
the convolution method. The molecule's zero-point energy has been chosen as the zero of the energy scale, and the calculation was made 
for bins with finite width of 1 cm$^{-1}$. To calculate the internal energy distribution of PAHs $P(E)$, we grouped this very large 
number of points into broader energy bins $[E^{min}_i,E^{max}_i]$ with $i=1$ to 500. Each energy bin $i$ thus contains many states and 
its degeneracy, $g_i$, is estimated as (with $g_{1} = 1$)
\begin{equation}
g_{i} = \int_{E^{min}_{i}}^{E^{max}_{i}} \rho(E) dE  \;\;\;\; \rm{(i} > 1\rm{)} \; .
\end{equation}

\section{Lines intensities in the rotational bands}
\label{AKJ}

In the case of symmetric top molecules, we give below the H\"onl-London factors for the angular part of rovibrational transition 
rates, as well as the corresponding transition energies \citep{Herzberg}. The frequency of the vibrational mode $i$ is noted $\nu_{i0}$.
\begin{itemize}
\item[-] for $\Delta J = 0, \pm 1$ and $\Delta K = 0$ (parallel transitions or out of plane transitions):
\begin{eqnarray}
A_{KJ^{+}} &=& \frac{2\left( (J+1)^{2}-K^{2} \right)}{(J+1)(2J+1)^2}  \;\;\; {\rm and} \;\;\; \nu_i^{+} = \nu_{i0}-2B(J+1) \nonumber \\
A_{KJ^{-}} &=& \frac{2(J^{2}-K^{2})}{J(2J+1)^2}  \;\;\;\;\;\;\;\;\;\;\;\; {\rm and} \;\;\;  \nu_i^{-} = \nu_{i0}+2BJ \nonumber  \\
A_{KJ^{0}} &=& \frac{2K^{2}}{J(J+1)(2J+1)}   \;\;\;\; {\rm and} \;\;\; \nu_i^0 = \nu_{i0} \nonumber  \\
\end{eqnarray}
\item[-] for $\Delta J = 0, \pm 1$ and $\Delta K = \pm 1$ (perpendicular or in plane transitions):
$$
\left\{
\begin{array}{l}
\displaystyle
A_{KJ^{+}} = \frac{(J+2\pm K)(J+1\pm K)}{2(J+1)(2J+1)^2}\\
\\
\nu_i^{+} = \nu_{i0}-2B(J+1)+(B-C)(1 \pm 2K)\\
\end{array}
\right.
$$
$$
\left\{
\begin{array}{lcl}
\displaystyle
A_{KJ^{-}} = \frac{(J-1 \mp K)(J\mp K)}{2J(2J+1)^2}\\
\\
\nu_i^{-} = \nu_{i0}+2BJ+(B-C)(1 \pm 2K)\\
\end{array}
\right.
$$
$$
\left\{
\begin{array}{lcl}
\displaystyle
A_{KJ^{0}} = \frac{(J+1\pm K)(J\mp K)}{2J(J+1)(2J+1)}\\
\\
\nu_i^{0} = \nu_{i0}+(B-C)(1 \pm 2K)\\
\end{array}
\right.
$$
\end{itemize}

\section{Rate of angular momentum change for gas-grain interactions}
\label{gas_interaction}

We describe below the gas-grain interactions considered in our model of PAH rotation and the rate of change of $J$
they induce, $(\tau^{-1}\Delta J)$. 

\subsection{Collisions with gas atoms and plasma drag}

For the collisions with gas neutrals and ions, as well as the plasma drag, we apply the results of DL98 to planar PAHs.
In the case where $J>>1$, we use the correspondence principle to write $\hbar J = I_c\omega$ and obtain the following rates:
\begin{eqnarray}
(\tau^{-1}\Delta J) = - 5.2\times 10^{-11} \sqrt{\frac{T}{100\,\rm{K}}} \left(\frac{n_{H}}{100\,\rm{cm}^{-3}}\right)\times J \times F
\end{eqnarray}
for the damping contribution, and
\begin{eqnarray}
(\tau^{-1}\Delta J) = 3.9\times 10^{-6}\left(\frac{T}{100\,\rm{K}}\right)^{\frac{3}{2}}\left(\frac{n_{H}}{100\,\rm{cm}^{-3}}\right)\left(\frac{N_{C}}{50}\right)^{2}
\times\frac{1}{J}\times G
\end{eqnarray}
for the exciting contribution, where $F$ and $G$ are normalized rates defined in the Appendix B of DL98. We use the 
formalism of \citet{Bakes} to estimate the average charge of PAHs of a given size.

\subsection{Rocket effect}

Ejection of H or H$_2$ from the edges of PAHs may yield a significant rotational excitation if it occurs asymmetrically, thus 
generating a systematic torque by the rocket effect \citep{Rouan92}. We note below $E_{ej}$ the kinetic energy of the ejected 
fragment. In the case of H$_2$, this will happen if this molecule forms on preferential sites by chemisorption, and if the 
distribution of these sites on the PAH is asymmetric as a result of dehydrogenation. We calculate the change of $J$ assuming 
that H$_{2}$ molecules are ejected from the edge of the PAH with a cosine law:
\begin{eqnarray}
(\tau^{-1} \Delta J)_{H_{2}} = 3.4 \times 10^{-8} \; \left(\frac{N_{C}}{50}\right)^{3/2} \; \left(\frac{n_{H}}{100 \ \rm{cm}^{-3}}\right)\nonumber\\
\times \;\sqrt{\frac{T}{100\ \rm{K}}} \; \sqrt{\frac{E_{ej}}{1.5\ \rm{eV}}} \; (1-f_{\rm H}) \; \; \rm{s}^{-1} \; .
\end{eqnarray}
The above numerical values are based on the following assumptions: (i) all molecular hydrogen is formed on PAHs with $[C/H]_{PAH} 
= 4\times10^{-5}$, a H$_{2}$ formation rate $R_{f} = 3 \times 10^{-17} (T/70 \rm{K})^{1/2}$~cm$^{3}$s$^{-1}$ \citep{Jura1975} 
and $E_{ej} = 1.5$ eV ; (ii) the distribution of formation sites has an asymmetry of 1 site and we assume that the site in excess 
is always at the same location on the molecule ; (iii) we neglect the influence of cross-over events that may reduce the angular 
momentum \citep{Lazarian1999}. With all these assumptions, the spin rate due to H$_{2}$ formation estimated here is an upper limit. 
This rate, however, remains small compared to the other gas-grains processes.

\subsection{Photoelectric effect}

Stellar UV photons can pull out electrons from grains. These photoelectrons carry away a significant kinetic 
energy ($\sim 1$eV) that heats the interstellar gas and impulses grain rotation. If we assume that the photoelectrons leave the 
grain surface with a cosine law distribution, we have
\begin{eqnarray}
(\tau^{-1} \Delta J)_{pe} = \tau_{pe}^{-1} \times 0.15 \ \sqrt{\frac{N_{C}}{50}} \ \sqrt{\frac{E_{e-}}{1\ \rm{eV}}} \; \; \ \ \ \rm{s}^{-1}
\end{eqnarray}
where $\tau_{pe}^{-1}$ is the rate of photoelectrons ejections calculated with the formalism of \citet{Bakes} ($\tau_{pe}^{-1} = 
1.6 \times 10^{-8}$ s$^{-1}$ for $N_{C} = 50$ in the DIM).

\section{Rotational temperature}
\label{temperature_rotationnelle}

\begin{figure}[!t]
\centering
\resizebox{\hsize}{!}{\includegraphics{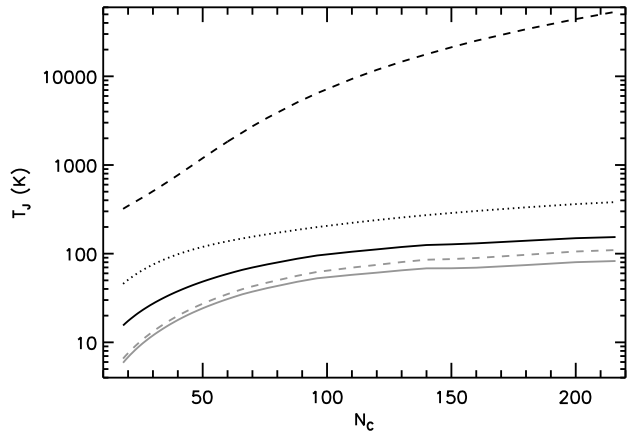}}
\caption{Rotational temperature for PAH cations as a function of their size for several insterstellar environments: CNM (black line), WNM (gray 
dashed line), WIM (gray line), MC (black dotted line) and the Orion Bar (black dashed line).}
\label{figure_Trot}
\end{figure}

A rotational temperature $T_J$ can be defined from the Maxwell distribution of angular momentum, $n(J)=n_0\exp(-J^2/J_0^2)$:
\begin{equation}
T_J({\rm K}) = \frac{BJ_0^2}{k_B} = 10 \times N_C^{-2}J_0^2 \; .
\end{equation}
Figure \ref{figure_Trot} shows this temperature as a function of PAH size for several interstellar environments. We see that $T_J$ is 
subthermal in the case of the DIM, and suprathermal for MCs and the Orion Bar. These results are in good agreement with \citet{Rouan97}. 
The rotational temperature is a relevant parameter in the study of the width of PAH vibrational or electronic transitions. The latter have been proposed 
as the origin of some unidentified diffuse interstellar bands (DIBs).

\end{appendix}

\end{document}